\documentclass[aps,prl,showpacs,twocolumn,superscriptaddress,letterpaper,longbibliography]{revtex4-1}
\usepackage{graphicx}
\usepackage{dcolumn}
\usepackage{bm}
\usepackage{amsmath, amssymb}%
\usepackage{mathtools}
\usepackage{epstopdf}
\usepackage{color}
\usepackage{soul}
\usepackage{hhline}
\usepackage{multirow}
\usepackage[usestackEOL]{stackengine}
\usepackage{placeins}
\usepackage{braket}
\usepackage{lineno}

\usepackage{hyperref}  
\hypersetup{breaklinks = true, colorlinks = true, citecolor = blue, linkcolor = blue, urlcolor = blue}
\usepackage{cleveref}
\graphicspath{{figures/}}

\setstackgap{L}{\normalbaselineskip}

\newcommand{\prel}{\ifmmode p_{rel} \else $p_{rel}$ \fi}
\newcommand{\pcm}{\ifmmode p_{cm} \else $p_{cm}$ \fi}

\newcommand{\mom}{{\bf p}}

\usepackage[]{changes} 

\begin{document}

\title{Short-Range Correlations and the Nuclear EMC Effect in Deuterium and Helium-3}

\newcommand*{\MIT }{Massachusetts Institute of Technology, Cambridge, Massachusetts 02139, USA}
\newcommand*{\StonyBrook}{Stony Brook, State University of New York, NY}
\newcommand*{\ODU}{Old Dominion University, Norfolk, Virginia 23529}
\newcommand*{\JLAB}{Thomas Jefferson National Accelerator Facility, Newport News, Virginia 23606}
\newcommand*{\TAU }{School of Physics and Astronomy, Tel Aviv University, Tel Aviv 69978, Israel}
\newcommand*{\Penn}{Pennsylvania State University, University Park, PA, 16802}
\newcommand*{\UW}{Department of Physics, University of Washington, Seattle, WA 98195-1560, USA}
\newcommand*{\GW}{George Washington University, Washington, DC 20052, USA}

\author{E.P.~Segarra}
\affiliation{\MIT}
\author{J.R. Pybus}
\affiliation{\MIT}
\author{F.~Hauenstein}
\affiliation{\MIT}
\affiliation{\ODU}
\author{D.W. Higinbotham}
\affiliation{\JLAB}
\author{G.A. Miller}
\affiliation{\UW}
\author{E.~Piasetzky}
\affiliation{\TAU}
\author{A. Schmidt}
\affiliation{\GW}
\author{M. Strikman}
\affiliation{\Penn}
\author{L.B. Weinstein}
\affiliation{\ODU}
\author{O. Hen}
\email[Contact Author \ ]{hen@mit.edu}
\affiliation{\MIT}

\date{\today{}}

\begin{abstract}
The EMC effect in deuterium and helium-3 is studied using a convolution formalism that allows isolating the impact of high-momentum nucleons in short-ranged correlated (SRC) pairs. 
We assume that the modification of the structure function of bound nucleons  is given by a universal (i.e. nucleus independent) function of their virtuality,
and find that the effect of such modifications is dominated by nucleons in SRC pairs.
This SRC-dominance of nucleon modifications is observed despite the fact that the bulk of the nuclear inelastic scattering cross-section comes from interacting with low-momentum nucleons. These findings are found to be robust to model details including nucleon modification function parametrization, free nucleon structure function and treatment of nucleon motion effects. While existing data cannot discriminate between such model details, we present predictions for measured, but not yet published, tritium EMC effect and tagged nucleon structure functions in deuterium that are sensitive to the neutron structure functions and bound nucleon modification functions.
\end{abstract}

\maketitle

\section{Introduction}
Determining the underlying cause of the modification of the partonic structure of nucleons bound in atomic nuclei, 
known as the EMC effect~\cite{Arnold:1984,Aubert83,Ashman88,Gomez94,Arneodo90,Seely09,Schmookler:2019nvf}, is an outstanding question in nuclear physics. 
Decades after its discovery, there is still no universally accepted explanation for the origin of the EMC effect~\cite{Frankfurt88,Norton03,Hen:2016kwk},
despite a large number of high-precision measurements in a wide variety of atomic nuclei. 

Modern models of the EMC effect account for both `conventional'
nuclear physics effects such as Fermi-motion and binding, as well as
for the more `exotic' effects of nucleon
modification~\cite{Norton03,Hen:2016kwk}.  The conventional nuclear
physics effects are well understood and cannot reproduce experimental
data alone, especially when including Drell-Yan
data~\cite{Alde:1990im,Hen:2016kwk}.  While required to reproduce
experimental data, nucleon modification models are far less
constrained and their microscopic origin is 
debated~\cite{Hen:2016kwk}.

An observed correlation between the magnitude of the EMC effect and the relative amount 
of short-range correlated (SRC) nucleon pairs in different nuclei~\cite{weinstein11,Hen12,Hen:2013oha,Schmookler:2019nvf} suggests that the EMC effect is driven by the modification of nucleons in SRC pairs. SRCs are pairs of strongly interacting nucleons at short distances. Nucleons in SRC pairs have large spatial overlap between their quark distributions and are highly offshell ($E^2\neq |{\bf{p}}|^2+m^2$), which makes them prime candidates for structure modification. 

Most recently, it has been demonstrated~\cite{Schmookler:2019nvf,Segarra:2019gbp} that the EMC effect in nuclei from helium-3 ($^3$He) to lead can be explained by a  single effective universal modification function (UMF) of nucleons in SRC pairs.
The UMF was constructed to be as model-independent as possible.
It is insensitive to the largely-unknown free-neutron structure function, $F_2^n$,
and accounts for both conventional nuclear effects, such as the scheme dependence of the deuteron wave-function, and nucleon motion effects, as well as more exotic nucleon modification effects.

Here we study the EMC effect using a convolution formalism that allows
us to separate the mean field and short range correlation
contributions of nucleon modification effects to the total UMF.  We
consider only light nuclei (the deuteron and $^3$He), for which exact
nuclear wave functions are available, and nucleon
modification effects can be isolated.  The sensitivity of the convolution
formalism to parametrization of the nucleon modification function,
$F_2^n$, and the treatment of nucleon motion effects are studied.

We find that, as expected, the bulk of the structure-function comes
from interactions with low-momentum nucleons.  However, nucleon
modification effects, which are required for a complete reproduction
of the measured data, are dominated by nucleons in SRC pairs.  We also
find that existing data cannot discriminate between different $F_2^n$
models or different parameterizations of bound nucleon modification
functions.  We predict new observables that can constrain
these model inputs, including the tritium EMC effect, sensitive to
$F_2^n$, and deuterium tagged nucleon structure functions,
sensitive to bound nucleon modification functions.  These predictions will soon be
tested by data from the MARATHON~\cite{MARATHON},
BAND~\cite{band-proposal}, and LAD~\cite{Emcsrcexpt11} Collaborations.


\section{Formalism}
\subsection{$F_2^A$ Convolution Approximation}
\begin{figure}[t]
		\includegraphics[width=0.49\textwidth]{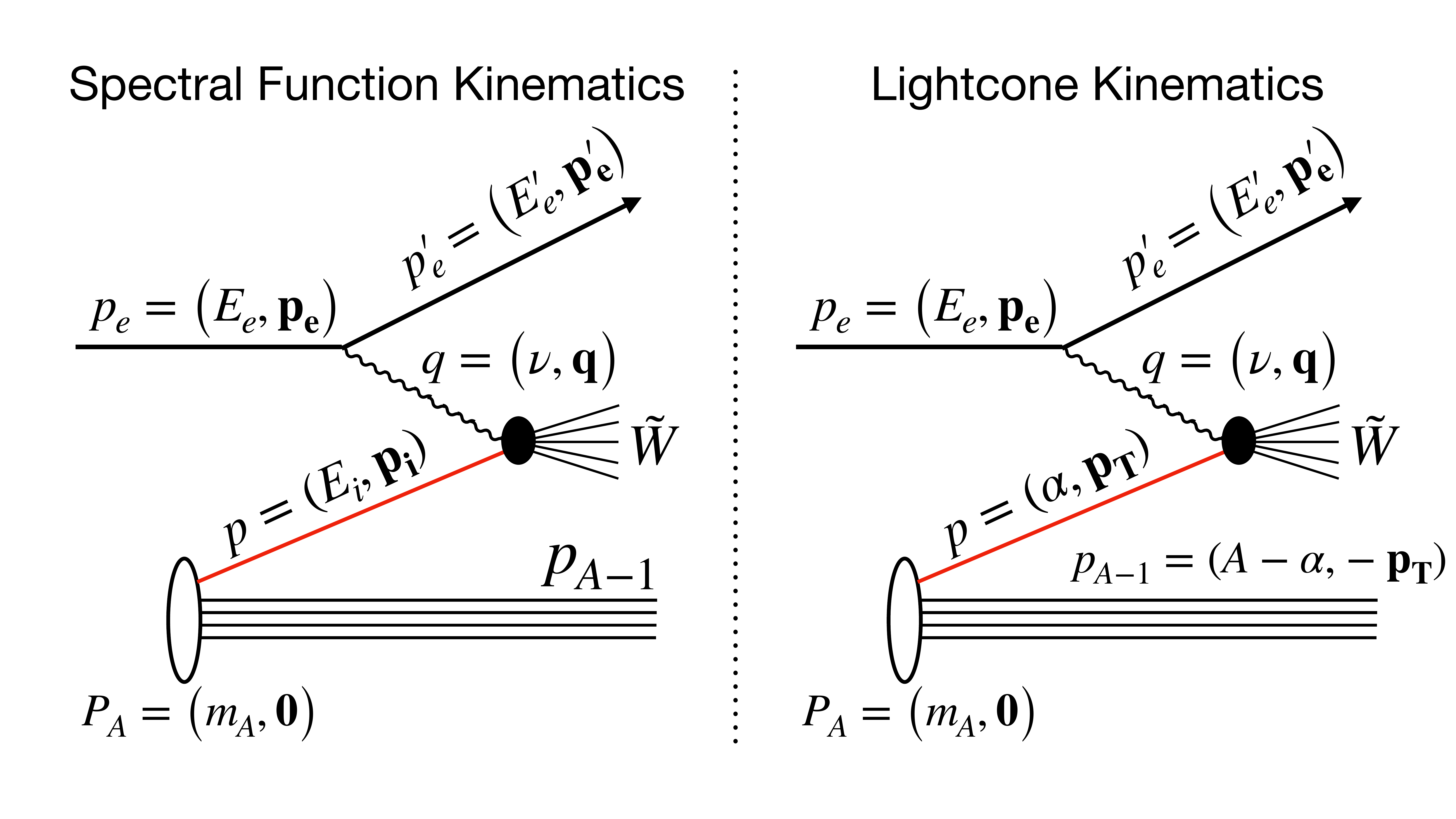}
		\caption{Reaction diagram for lepton-nuclear deep inelastic scattering in a factorized plane wave impulse approximation. Red lines represent off-shell nucleons. See text for details. }
		\label{fig:scatteringDiagram}
\end{figure}

In order to study the EMC effect in a framework that allows us to understand its dependence on nucleon momentum and offshellness, we calculate the nuclear structure function, $F_2^A(x_B)$, using the nuclear convolution model for lepton-nucleus DIS~\cite{Frankfurt:1985cv,FSemc:1987,Frankfurt88,SAKulaginSVAku:1985,DunneThomas:1985}:
\begin{small}
\begin{equation}
\begin{split}
		&F_2^A(x_B) = \\
		& \frac{1}{A} \int_{x_B}^A \frac{d\alpha}{\alpha} \int_{-\infty}^0 dv \Big[Z \tilde{\rho}_p^A(\alpha,v) F_2^p\big(\tilde{x}\big) + N \tilde{\rho}_n^A(\alpha,v) F_2^n\big(\tilde{x}\big) \Big] \\
		& \;\;\;\;\;\;\;\;\;\;\;\;\;\;\;\; \times \Big(1+v \ f^{off}(\tilde{x})\Big)  \\
		& = \frac{1}{A}\int_{x_B}^A \frac{d\alpha}{\alpha} \int_{-\infty}^0 dv \: F_2^p\big(\tilde{x}\big) \left[ Z \tilde{\rho}_p^A(\alpha,v) + 
		N \tilde{\rho}_n^A(\alpha,v) \frac{F_2^n(\tilde{x})}{F_2^p(\tilde{x})} \right]   \\
		& \;\;\;\;\;\;\;\;\;\;\;\;\;\;\;\;  \times  \Big(1+v\ f^{off}(\tilde{x})\Big),
\label{Eq:F2_LC}
\end{split}
\end{equation}
\end{small}
where $x_B=Q^2/(2m_N\nu)$, $Q^2$ is the four momentum transfer
squared, $m_N$ is the nucleon mass and $\nu$ is the energy transfer
(Fig.~\ref{fig:scatteringDiagram}).  $\tilde{x}=\frac{Q^2}{2 p\cdot
  q}$ where $q$ is the four-momentum of the virtual photon and $p$ is
the initial four-momentum of the struck off-shell nucleon. $\tilde{x}$ reduces
to $\frac{x_B}{\alpha}\frac{m_N A}{m_A}$ in the Bjorken limit with
lightcone momentum fraction $\alpha = A(E+p_z)/m_A$ (see online
supplementary materials for finite energy corrections to
Eq.~\ref{Eq:F2_LC} at low $Q^2$).  Here $z$ is opposite to the direction
of the virtual photon, and $v = (E^2-|{\bf{p}}|^2 -
m_N^2)/m_N^2$ is the bound nucleon fractional virtuality.
$\tilde{\rho}_N^A(\alpha,v)$ are the nucleon ($N$ = $p$ or $n$)
lightcone momentum and virtuality distributions in nucleus $A$,
defined below.  $F_2^p \big(\tilde{x}\big)$ and $F_2^n
\big(\tilde{x}\big)$ are the free proton and neutron structure
functions. For brevity we omit their explicit $Q^2$ dependences but
note that $F_2^p$, $F_2^n$, and $F_2^A$ are always evaluated at the
same $Q^2$ value.  $f^{off}(\tilde{x})$ is a universal offshell
nucleon modification function, assumed here to be the same for
neutrons and protons and for all nuclei. In Eq.~\ref{Eq:F2_LC}, we
take the offshell effect to be linear in $v$ (i.e. $1+v
f^{off}(\tilde{x})$ ) as a first-order Taylor expansion in
virtuality; see Ref.~\cite{Miller:2019mae} for additional discussion.

\subsection{Lightcone densities}
In our convolution, traditional nuclear contributions to the EMC
effect such as nucleon motion and binding are treated within the
one-body lightcone momentum and virtuality distribution,
$\tilde{\rho}^A_{N}(\alpha,v)$.  It describes the joint
probability to find a nucleon ($n$ or $p$) in a nucleus $A$ with
lightcone momentum fraction $\alpha$ and fractional virtuality
$v$. Integrating over fractional virtuality defines the
lightcone momentum distribution of a nucleon
\begin{equation}
\rho_N^A(\alpha) =  \int_{-\infty}^0 dv \tilde{\rho}_N^A(\alpha,v),
\end{equation}
that is normalized herein according to the baryon sum rule:
\begin{equation}
 \int_{0}^A \frac{d\alpha}{\alpha}   {\rho}_N^A(\alpha) \equiv 1.
\label{Eq:baryon_SR}
\end{equation}

To avoid producing an artificial EMC like effect in nucleon-only models when used in Eq.~\ref{Eq:F2_LC}~\cite{FSemc:1987}, ${\rho}_N^A(\alpha)$ must also satisfy the momentum sum rule:
\begin{equation}
\frac{1}{A}\int_{0}^A \frac{d\alpha}{\alpha}  \alpha \left( Z {\rho}_p^A(\alpha) + N {\rho}_n^A(\alpha) \right) = 1.
\label{Eq:momentum_SR}
\end{equation}

It is necessary to know the functional form of $\tilde\rho_N^A(\alpha,v)$ to proceed further. Although the nuclear wave functions for nuclei with  $A=2$ and $A=3$ have been well-computed, they do not suffice to unambiguously yield the light-cone momentum distributions and their dependence on virtuality. This is because current calculations are non-relativistic and made with an underlying assumption that the nucleons are on their mass shell. Handling this issue on a fundamental level would require a first-principles light-front calculation including the effects of off-mass-shell dependence. Such a calculation could be done by solving the relevant Bethe-Salpeter equation, but does not yet exist. 

Therefore, we consider here two approximations to estimate $\tilde{{\rho}}(\alpha,\nu)$: a spectral-function (SF) approximation, 
where the momentum sum rule is violated if only nucleonic degrees of freedom are taken into account, and a generalized-contact formalism lightcone (GCF-LC) approximation.

\begin{figure*}[ht!]
		\includegraphics[width=0.32\textwidth]{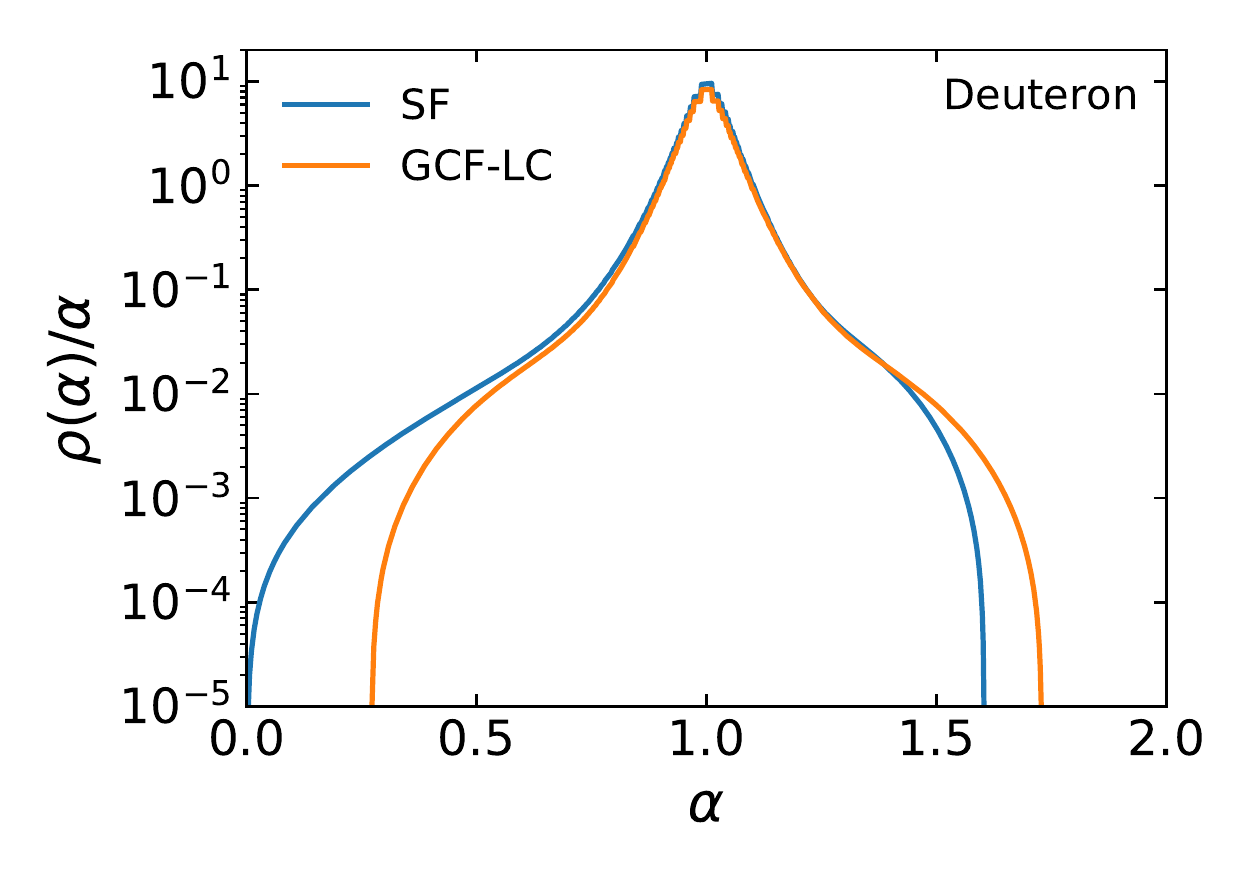}
		\includegraphics[width=0.32\textwidth]{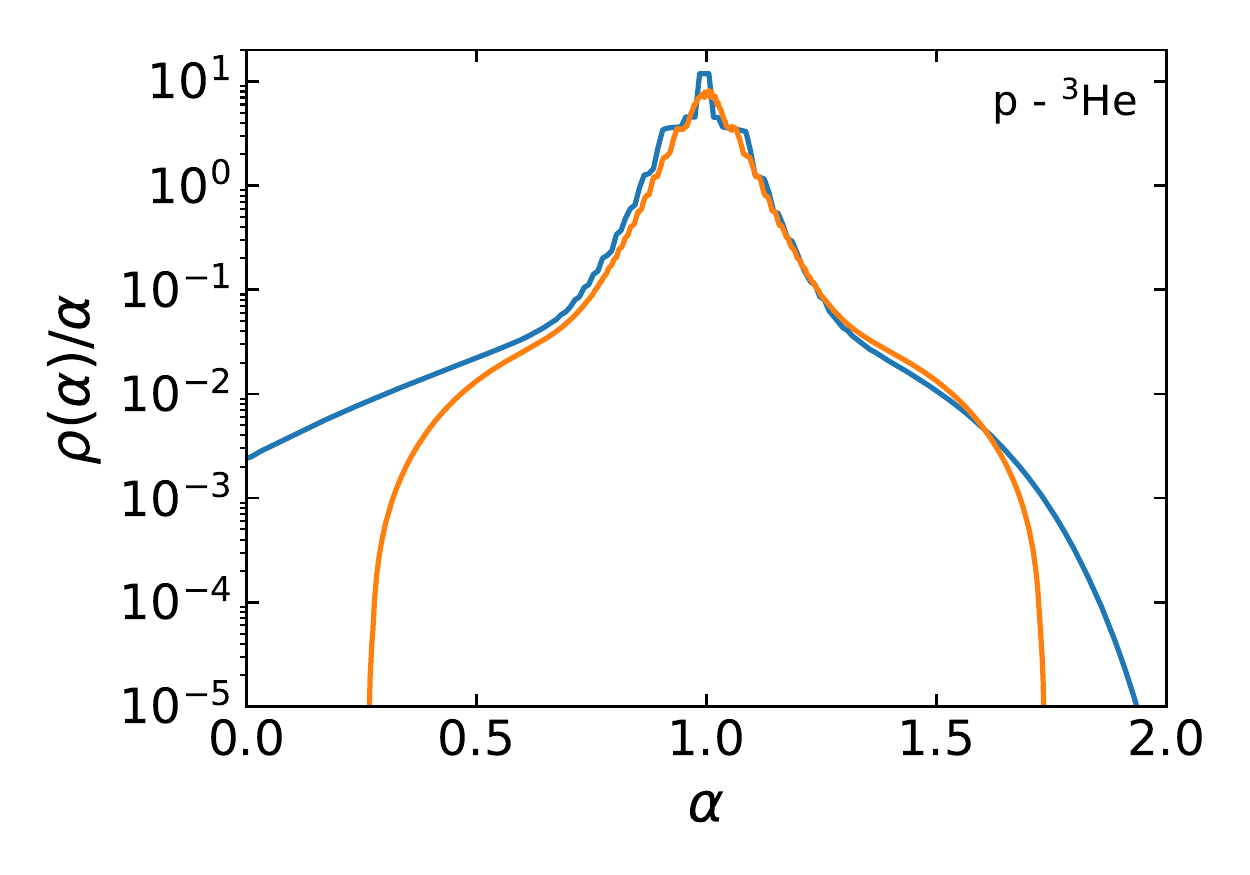}
		\includegraphics[width=0.32\textwidth]{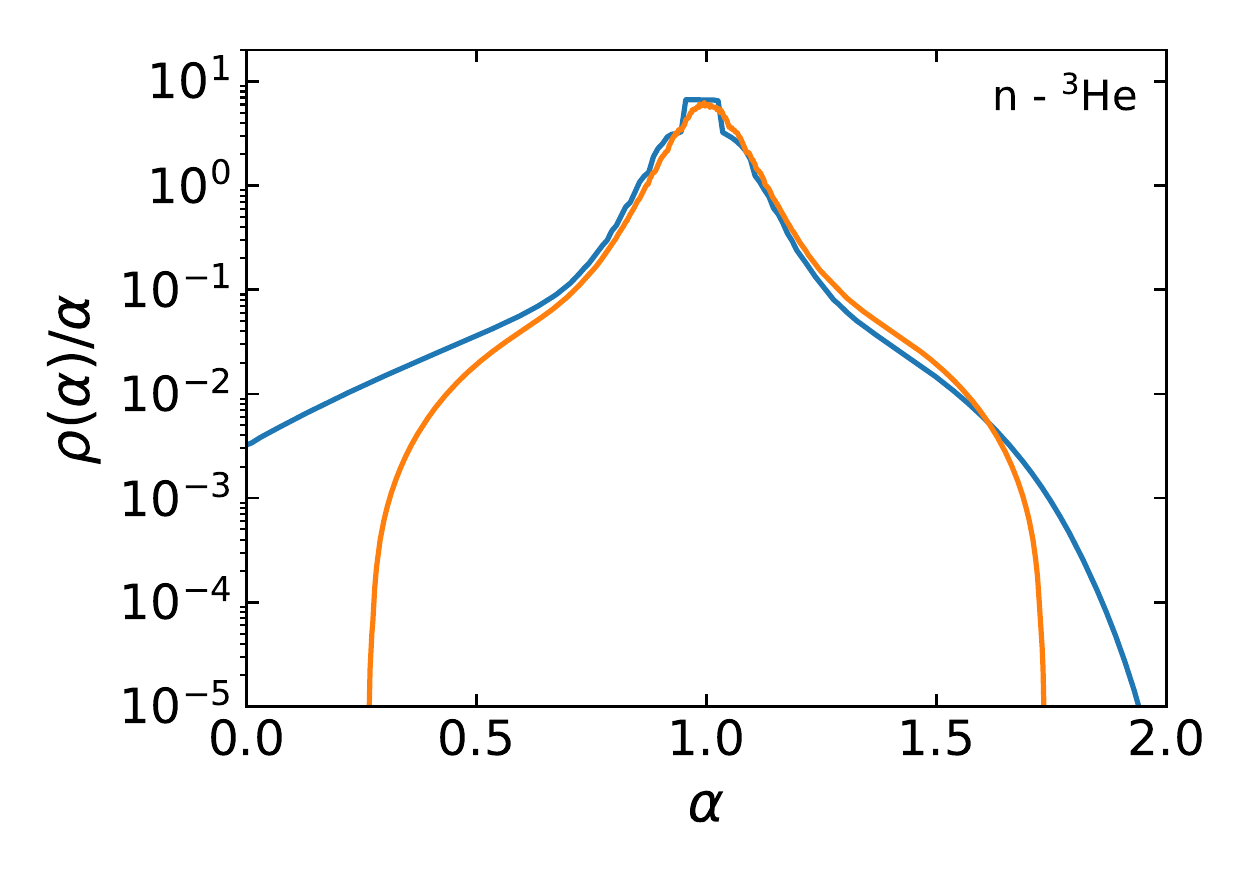}\\
		\caption{Lightcone momentum distributions $\rho(\alpha)$ for deuteron (left) and protons (center) and neutrons (right) in $^3$He calculated using the spectral function (SF) and generalized contact formalism lightcone (GCF-LC) approximations. The discretization visible in the SF distributions (blue lines) is due to the discretization of the spectral function $S(E,p)$ and integration of Eq.~\ref{Eq:LC_SF}.}
		\label{fig:alpha_LC_SF}
\end{figure*}

\subsubsection{Spectral function approximation}
The nuclear spectral function $S(E,p)$ defines the probability for finding a nucleon in the nucleus with momentum $p$ and nucleon energy $E$.
Exactly calculable spectral functions are available for light nuclei and allow calculating the nuclear lightcone distributions as~\cite{Miller:1990,Miller:1988}
\begin{small}
\begin{equation}
\begin{split}
\tilde{\rho}_{N, SF}^A(\alpha,v) &=  \int dE d^3\mom  \ \ S^A_N(E,p)   
					\cdot \frac{E+p_z}{E} \\ &\ \delta\Bigg(\alpha - \frac{A p^+}{P^+} \Bigg) \delta\Bigg(v - \frac{E^2-|{\bf{p}}|^2-m_N^2}{m_N^2} \Bigg),
\label{Eq:LC_SF}
\end{split}
\end{equation}
\end{small}
where $p={|\bf{p}|}$, 
$p^+\equiv E+p_z=m_A\alpha /A$ is the plus-component of the momentum of the struck nucleon, $P^+=m_A$ is the plus-component of the momentum of the nucleus $A$, and $m_A$ is the nucleus mass.

The flux factor $\left(E+p_z\right)$ is introduced to help satisfy the momentum sum rule~\cite{FSemc:1987}. The $\frac{1}{E}$ factor ensures SF-based lightcone distribution functions are appropriately normalized according to the Baryon sum-rule (Eq.~\ref{Eq:baryon_SR}). However, this also changes the interpretation of ${\rho}(\alpha)$ from a simple probability density for finding a nucleon in a nucleus with lightcone momentum fraction $\alpha$ (see discussion in refs.~\cite{FSemc:1987,Miller:1988,Miller:1990}).\\

For deuterium, considering a wave function calculated using the AV18 interaction, the momentum sum rule has a negligible violation ($<0.1\%$). For $^3$He, using the AV18-based spectral function of Ref.~\cite{AttiKaptari:2005}, it is violated by $\leq 1\%$. This small  violation is expected to produce an artificial EMC effect~\cite{FSemc:1987} that should result in a smaller nucleon modification effect required to explain the experimental data.

\subsubsection{Generalized-contact formalism lightcone approximation}
To fully satisfy the $^3$He momentum sum rule, we examine an alternative approach for calculating $\tilde \rho_N^A(\alpha,v)$ using a scale-separation approximation where the lightcone density function is separated into a mean-field (single-nucleon) part and an SRC part~\cite{Hen:2013oha,CiofidegliAtti:1995qe,Ciofi07,Weiss:2018tbu,Pybus:2020itv}:
\begin{small}
\begin{equation}
\begin{split}
\tilde{\rho}_{N, GCF-LC}^A (\alpha,v)=\tilde{\rho}_{N, GCF,SRC}^A (\alpha,v)+\tilde{\rho}_{N,MF}^A (\alpha,v).
\end{split}
\end{equation}
\end{small}

\begin{figure}[b]
		\includegraphics[width=0.4\textwidth]{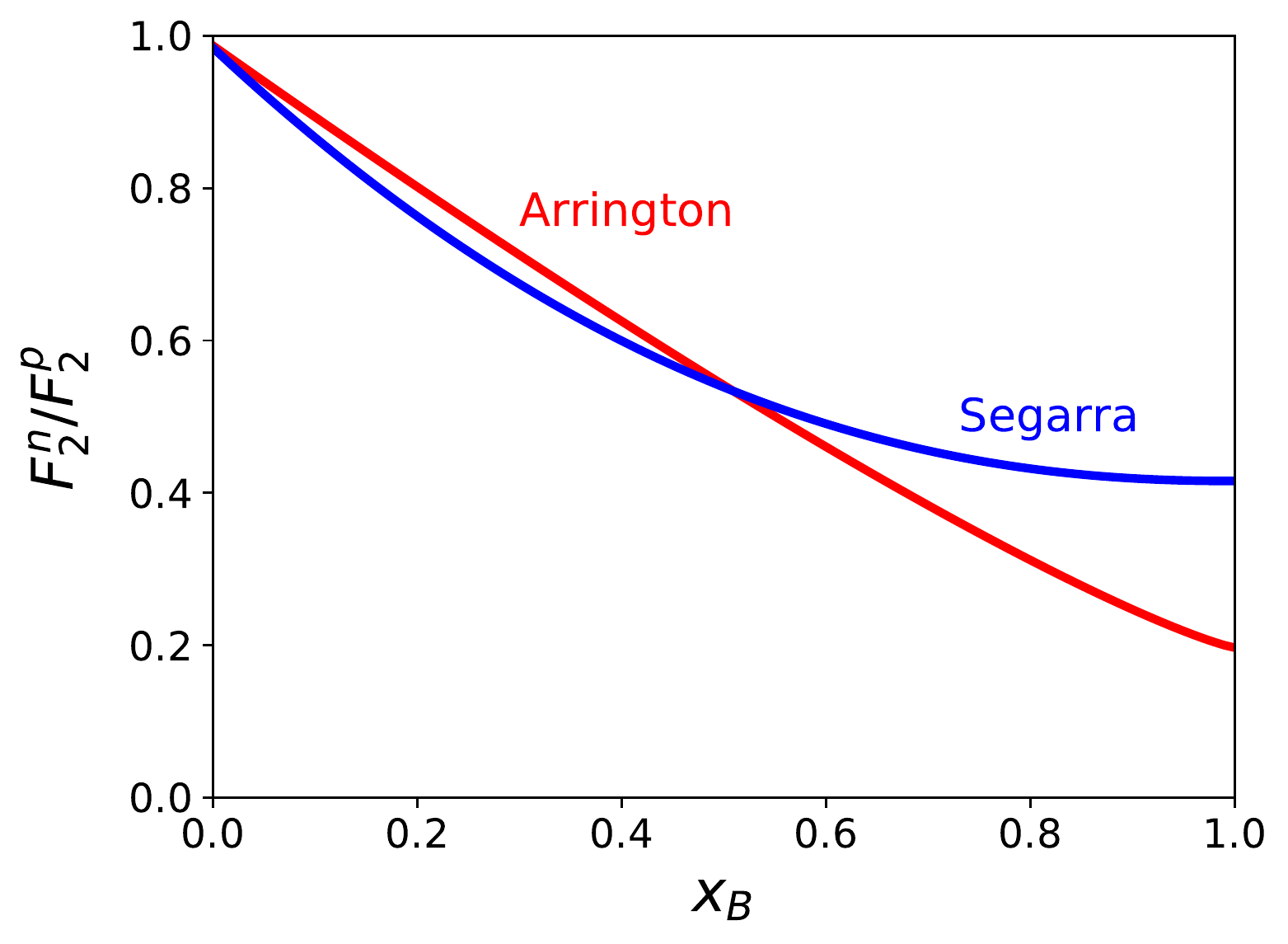}
		\caption{$F_2^n/F_2^p$ parametrizations used in this work that span the current range of models~\cite{Segarra:2019gbp}. See text for details.
		}
		\label{fig:F2nF2p}
\end{figure}

The SRC part of the lightcone density can be formulated by integrating over the lightcone SRC decay function~\cite{Pybus:2020itv,schmidt20}, which describes the distribution of the momentum of the struck nucleon as well as its partner, here denoted the `spectator' nucleon:
\begin{small}
\begin{equation}
\begin{split}
\tilde{\rho}_{N, GCF,SRC}^A (\alpha,v) = &\int d^2\mom_\perp \frac{d\alpha_s}{\alpha_s} d^2\mom_s^{\perp} \rho_{SRC}^N(\alpha,\mom^\perp,\alpha_s,\mom_s^{\perp})\\ 
&\times\delta\Bigg(v - \frac{p^-(m_A/A)\alpha - \mom_\perp^2-m_N^2}{m_N^2} \Bigg),
\end{split}
\end{equation}
\end{small}
where 
\begin{small}
\begin{equation}
\begin{split}
p^- &= P^- -  {p_s^-} - p_{A-2}^-\\
&= m_A - \frac{m_N^2+(\mom_s^\perp)^2}{(m_A/A) \alpha_s} - \frac{m_{A-2}^2+(\mom_{CM}^\perp)^2}{(m_A/A) (A-\alpha - \alpha_s)}
\end{split}
\end{equation}
\end{small}
is the off-mass shell minus-component of the struck nucleon's momentum, $\alpha_s$ is the spectator nucleon lightcone fraction, $\mom_\perp$ and $\mom_{s,\perp}$ are the transverse momentum of the struck nucleon and the spectator, respectively, and $\mom_{CM}^\perp=\mom_\perp+\mom_{s,\perp}$.
$\rho_{SRC}^N$ is a two-body (i.e. pair) lightcone density given by a convolution of the pair center-of-mass and relative  momentum densities, see Ref.~\cite{Pybus:2020itv} and online supplementary materials for details.

The mean-field part of the lightcone density is taken from the spectral functions using a linearized approximation, similar to Eq.~\ref{Eq:LC_SF}, but which manifestly preserves the baryon number  and momentum sum rules:
\begin{small}
\begin{equation}
\begin{split}
\tilde{\rho}_{N,MF}^A(\alpha,v)=\alpha\int_0^{m_N} dE \int_0^{p_{cutoff}}d^3\mom S_N^A(E,p)
\\ \times\delta\Bigg(\alpha - 1 - \frac{Ap_z}{P^+} \Bigg) \delta\Bigg(v- \frac{E^2-{|\bf{p}|}^2-m_N^2}{m_N^2} \Bigg).
\end{split}
\end{equation}
\end{small}
The cutoff momentum $p_{cutoff}=240$ MeV$/c$ for $^3$He and was chosen such that the fraction of SRC pairs was equal to that extracted from ab-initio many-body calculations ($10.1\%$ for neutrons and $5.9\%$ for protons)~\cite{Weiss:2016obx,Cruz-Torres:2019fum}.

We emphasize that the momentum sum rule for $\tilde{\rho}_{N, GCF-LC}^A(\alpha,v)$ is manifestly satisfied in this approximation
and that the resulting GCF-LC density is symmetric around unity, in contrast to that obtained in the SF approximation (Fig.~\ref{fig:alpha_LC_SF}).

\section{Structure Function and\\Modification Models}
We compute Eq.~\ref{Eq:F2_LC} using parameterizations of $f^{off}(\tilde{x})$, $F_2^p(\tilde{x})$, and  $\frac{F_2^n(\tilde{x})}{F_2^p(\tilde{x})}$, and both $\tilde{\rho}_{N,SF}^A$ and $\tilde{\rho}_{N,GCF-LC}^A$. For the modification function $f^{off}(\tilde{x})$ we consider three models:
\begin{eqnarray}
	f^{off}_{const}(\tilde{x}) &=& C, \\
	f^{off}_{lin\;x}(\tilde{x}) &=&a + b \cdot  \tilde{x}\\
	f^{off}_{KP,\;CJ}(\tilde{x}) &=& C(x_0-\tilde{x})(x_1-\tilde{x})(1+x_0-\tilde{x}) 
\end{eqnarray}
where $f^{off}_{const}$ assumes a virtuality-dependent modification
model that is independent of $\tilde{x}$, and $f^{off}_{lin\;x}$ is also
linearly dependent on $\tilde{x}$.  The  free parameters of these parameterizations ($C$, $a$, and $b$) are determined by fitting Eq.~\ref{Eq:F2_LC} to experimental data as detailed below.

We also use modification functions determined by KP ($f^{off}_{KP}$)~\cite{KULAGIN2006126} and CJ ($f^{off}_{CJ}$)~\cite{Accardi:2016qay}, who both chose to use a 3rd order polynomial in $\tilde{x}$, albeit with different parameters.
These are used here with their original parameters, extracted in Refs.~\cite{KULAGIN2006126,Accardi:2016qay}.

$\frac{F_2^n(\tilde{x})}{F_2^p(\tilde{x})}$ was parametrized as: 
\begin{equation}
\frac{F_2^n(\tilde{x})}{F_2^p(\tilde{x})}\equiv R_{np}(\tilde{x}) = a_{np} \left( 1- \tilde{x}\right)^{b_{np}} + c_{np},
\label{Eq:F2pn}
\end{equation}
where $R_{np}(\tilde{x}\rightarrow1) = c_{np}$.
We fix the $a_{np}$, $b_{np}$, and $c_{np}$ parameters by fitting
Eq.~\ref{Eq:F2pn} to one of two recent predictions by
Segarra~\cite{Segarra:2019gbp} and by
Arrington~\cite{Arrington:2011qt}, that represent two extreme models
that capture the spread of current models~\cite{Segarra:2019gbp} (see
Fig.~\ref{fig:F2nF2p}).  We further assume that
$R_{np}(\tilde{x})$ has negligible $Q^2$ dependence.  We
note that the original $f^{off}_{KP}$ and $f^{off}_{CJ}$ extractions
were done using $F_2^n/F_2^p$ that are respectively similar to the
Segarra and Arrington models used herein.

$F_2^p(x_B,Q^2)$ was taken from GD11-P~\cite{GD11P:2011}. As DIS data
are typically given in the form of $F_2^A/F_2^d$ ratios to minimize
higher twist effects, the only explicit $Q^2$ dependence we assume is
that of $F_2^p(x_B,Q^2)$, that is assumed to be negligible in the ratio
$F_2^n/F_2^p$.

We estimated the parameters of $f^{off}_{const}$ and
$f^{off}_{lin\;x}$ using a $\chi^2$-minimization inference from a
simultaneous fit to both $F_2^{^3He}/F_2^d$~\cite{Seely09} and
$F_2^d/(F_2^p+F_2^n)$~\cite{Griffioen:2015hxa} data for $0.17 \leq x_B
\leq 0.825$. While data for $F_2^{^3He}/F_2^d$ of~\cite{Seely09}
extends up to $x_B\sim0.9$, these high-$x_B$ data are at low invariant
mass, $W$.  Requiring $W>1.4\ \text{GeV}$ ($W^2>2\ \text{GeV}^2$) in
the fitting procedure limited the data to
$x_B\le 0.825$. However, we extrapolate our predictions up to $x_B
\sim0.95$ for use by future measurements, such as
MARATHON~\cite{MARATHON}.  Isoscalar corrections previously applied to
$F_2^{^3He}/F_2^d$ data were removed and the quoted experimental
normalization uncertainties of each data set were accounted for in the
fit. In the calculation of each data point, $F_2^p$ is evaluated at
the $Q^2$ value of the data.  We performed 16 inference trials for
different model
assumptions for $\tilde{\rho}(\alpha,v)$, $F_2^n/F_2^p$ and $f^{off}$ (see
Tab.~\ref{Tab:chi2}).

\begin{figure}[t]
		\includegraphics[width=0.48\textwidth]{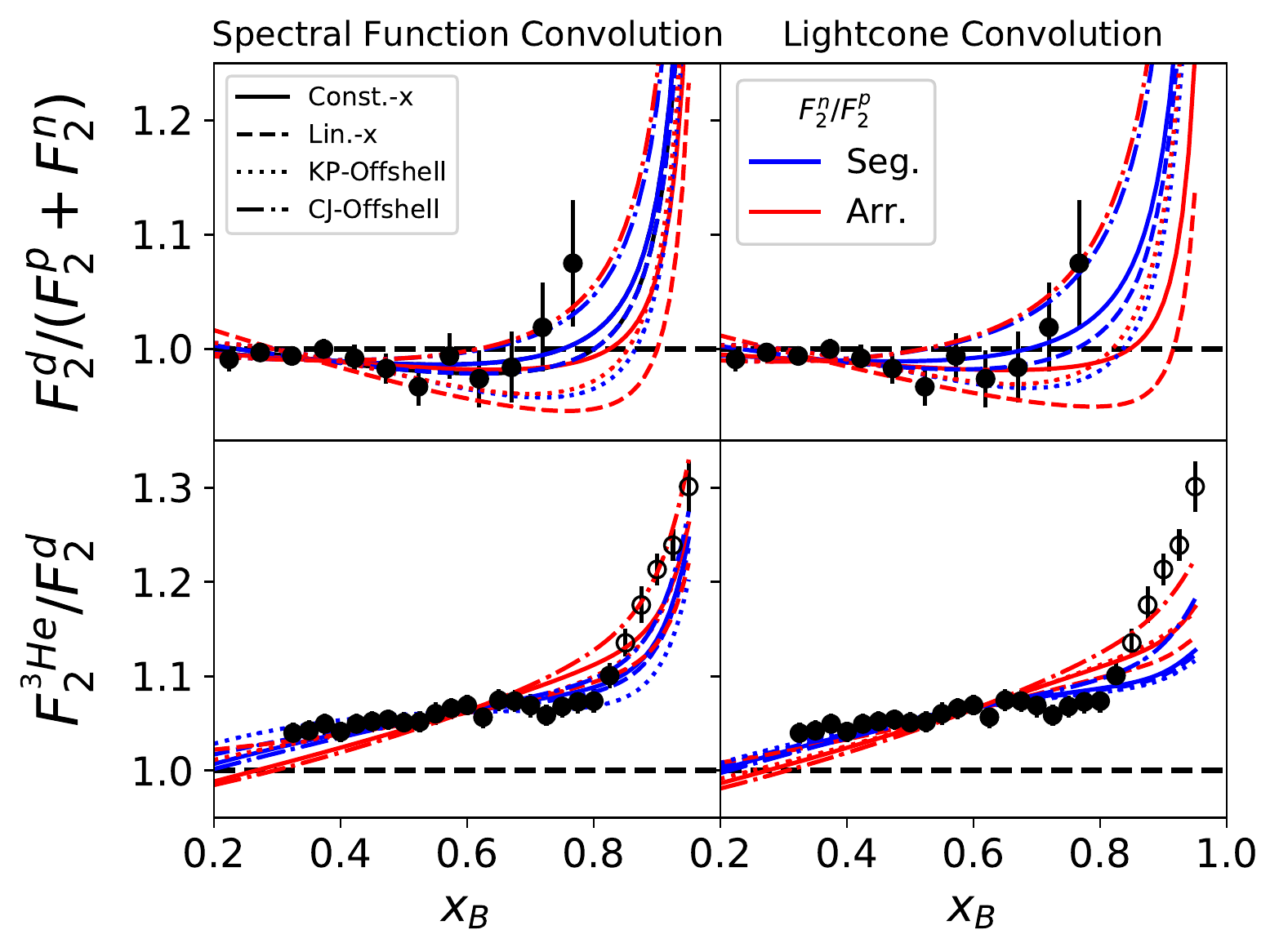}
		\caption{Convolution results after $\chi^2$-minimization procedure. Each model 
		is shown with the renormalization inferred for $^2$H and $^3$He and 
		the resulting $\chi^2$. Open circles denote data at $W<1.4$ GeV, which was not used in the fit.
		(Top-left) SF convolution results on $F_2^d/(F_2^p+F_2^n)$. 		
		(Top-right) LC convolution results on $F_2^d/(F_2^p+F_2^n)$.
		(Bottom-left) SF convolution results on $F_2^{^3He}/F_2^d$. 
		(Bottom-right) LC convolution results on $F_2^{^3He}/F_2^d$.
		All curves are calculated and extrapolated with the same $Q^2$ as the data for $^2$H and $^3$He.
		}
		\label{fig:ratios-fit}
\end{figure}

\section{Results}

\subsection{Inclusive data description}

\begin{figure}[t]
		\includegraphics[width=0.48\textwidth]{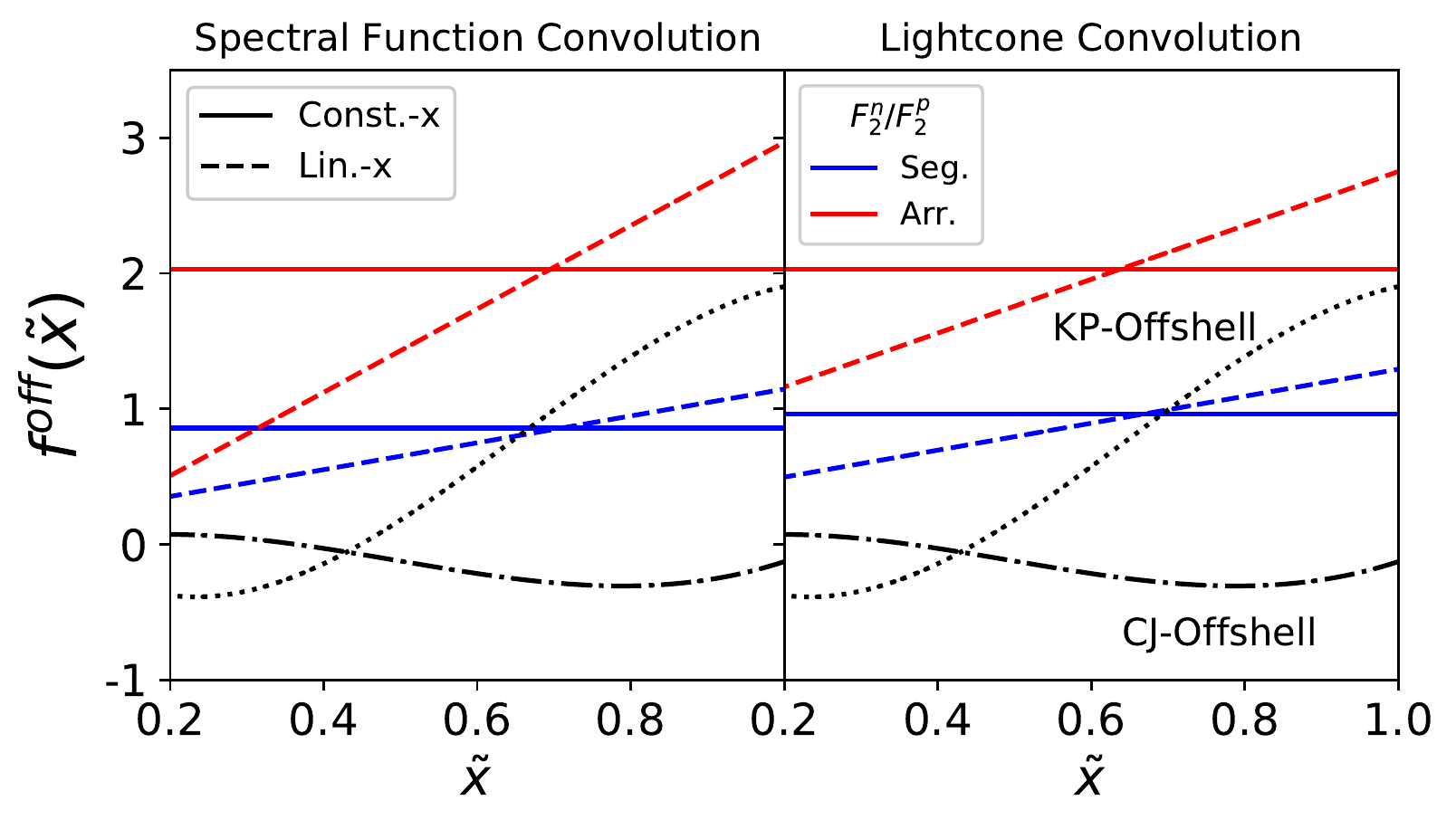}
		\caption{Offshell functions $f^{off}(\tilde{x})$ resulting from $\chi^2$-minimization procedure with SF (left) and LC (right) approximations. The blue and red curves were minimization trials using a $F_2^n/F_2^p$ fit to two recent predictions by Segarra~\cite{Segarra:2019gbp} (Seg.) and Arrington~\cite{Arrington:2011qt} (Arr.), respectively. The two black lines are the offshell functions as described in~\cite{KULAGIN2006126,Accardi:2016qay} and were taken as fixed for the minimization procedure which is why they are identical for both convolution frameworks. 
		}
		\label{fig:offshells}
\end{figure}

Figure~\ref{fig:ratios-fit} shows the resulting fit compared to the experimental data.
We performed eight individual fits, switching between the two $F_2^n/F_2^p$ models, 
constant-in-$\tilde{x}$ or linear-in-$\tilde{x}$ offshell parameterizations, and using either SF or GCF-LC lightcone densities. 
We also show calculations using $f^{off}_{KP,CJ}$ and both $F_2^n/F_2^p$ and lightcone densities.
For completeness, Fig.~\ref{fig:offshells} shows the inferred offshell
functions for $f^{off}_{const}$ and $f^{off}_{lin\;x}$ for the
different convolution frameworks, along with
$f^{off}_{KP,\;CJ}$~\cite{KULAGIN2006126,Accardi:2016qay}.

As can be seen from the strong overlap of many curves, 
these existing $^3$He and $^2$H data cannot definitely
discriminate between the different off-shell
function or nucleon motion effect models.  The data can be
adequately reproduced even with very different off-shell models.

Tab.~\ref{Tab:chi2} and Fig.~\ref{fig:ratios-fit} do show a systematic
improvement when using the Segarra et al. $F_2^n/F_2^p$
parametrization (blue curves). Using $f^{off}_{KP}$, the calculation
does not describe the high-$x_B$ $^2$H data. This is not unexpected as
their offshell function was not fit to BONUS data nor to high-$x_B$
deuterium data ($\ge 0.8$)~\cite{KULAGIN2006126}. $f^{off}_{KP}$ does
describe the $^3$He EMC data markedly well due to the global nature of
their analysis which captures the general EMC trend in a wide range of
nuclei. Similarly, when using $f^{off}_{CJ}$, the calculation
struggles as much as other models to accurately predict the $^3$He EMC
ratio. However, we note that their global fit does not consider $A>2$
nuclear DIS data. Again, the agreement improves with the use of the
Segarra et al. $F_2^n/F_2^p$.

\begin{figure}[t]
		\includegraphics[width=0.48\textwidth]{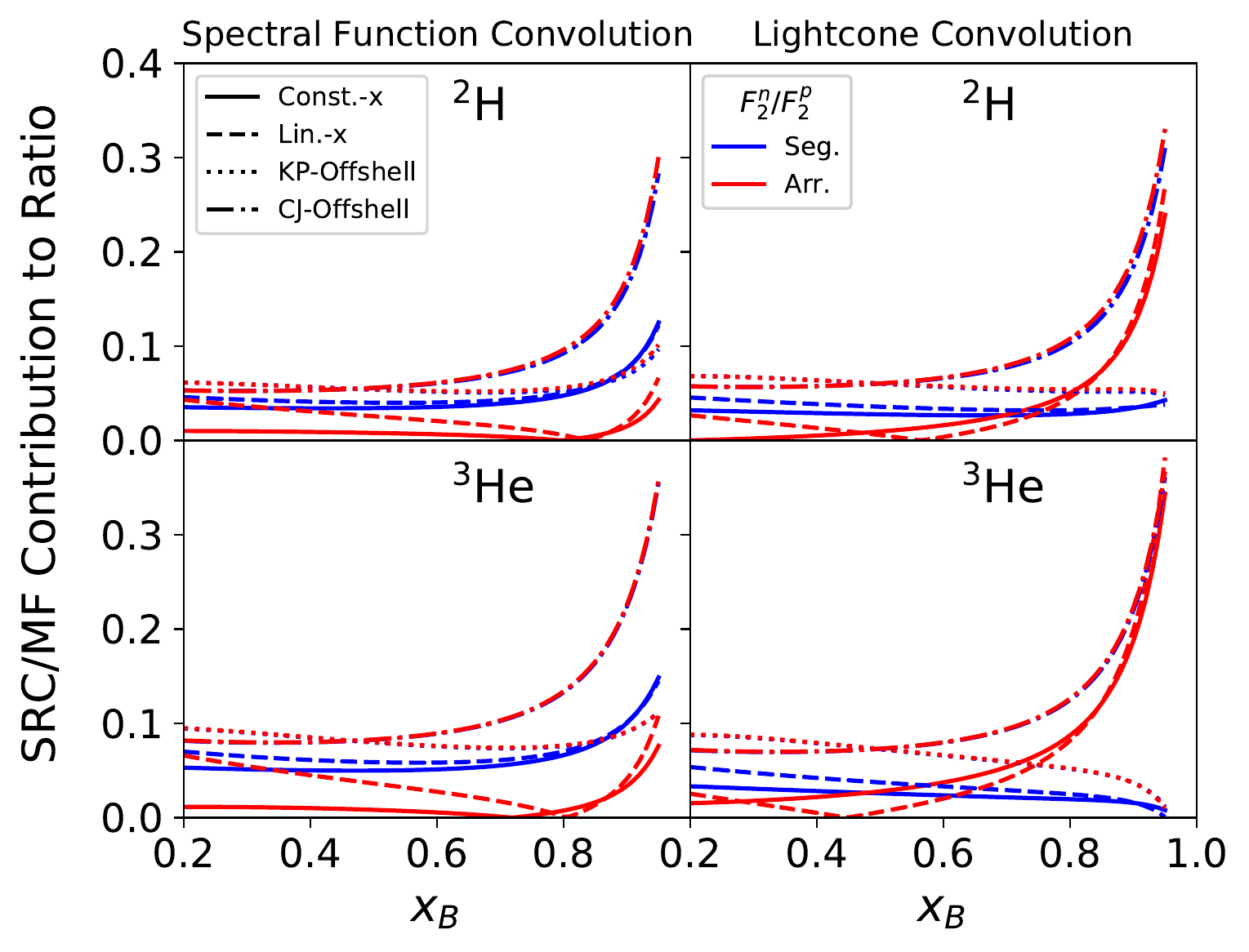}
		\caption{Ratio of SRC contribution to MF contribution of the structure function. 
		See text for details.
		(Top-left) SF ratio on $F_2^d$. 		
		(Top-right) LC ratio on $F_2^d$.
		(Bottom-left) SF ratio on $F_2^{^3He}$. 
		(Bottom-right) LC ratio on $F_2^{^3He}$.
		Curves are shown with $Q^2=5 \textrm{ GeV}^2$\\
		}
		\label{fig:ratios-breakdown}
\end{figure}

\begin{figure*}[t!]
		\includegraphics[width=0.914\textwidth]{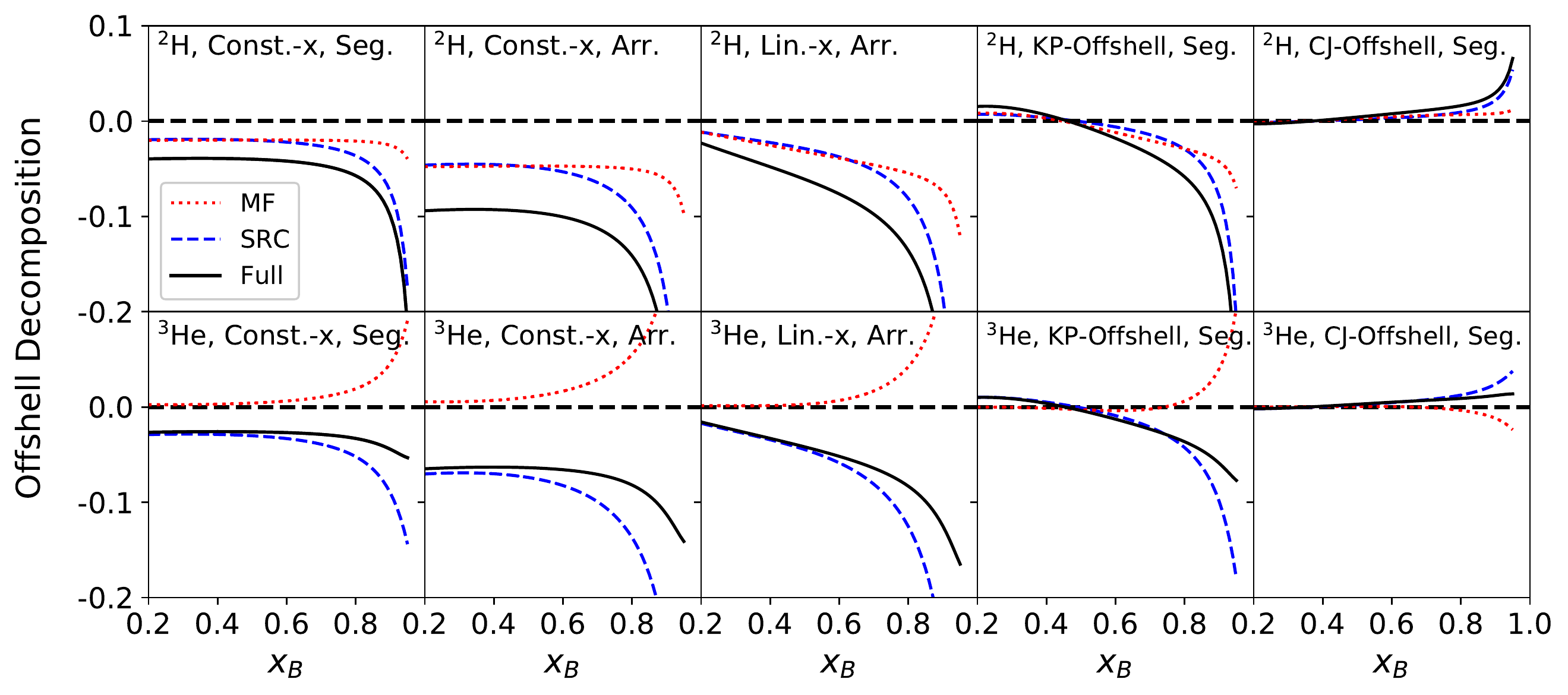}
		\centering	
		\caption{Decomposition of $F_2^A(\text{offshell})$ for various model assumptions within the SF approximation. Top: offshell effect in $^2$H for model 
		assumptions of (from left-to right) 1) constant-in-x with Seg. $F_2^n/F_2^p$, 2) constant-in-x with 
		Arr. $F_2^n/F_2^p$, 3) linear-in-x with Arr. $F_2^n/F_2^p$, 4) KP-offshell function with Seg. 
		$F_2^n/F_2^p$, and 5)  CJ-offshell function with Seg. $F_2^n/F_2^p$. Bottom: offshell effect 
		in $^3$He for the same models.
		Solid black lines represent the full offshell contribution. Dashed blue lines are the contribution due 
		to SRC nucleons ($>240$ MeV/c in the SF assumption). Similarly, dotted red lines are 
		the contribution due to MF nucleons ($<240$ MeV/c). Curves not shown for other model assumptions 
		considered can be viewed in the online supplementary materials. Curves are shown at 
		$Q^2=5 \textrm{ GeV}^2$.\\
		}
		\label{fig:offshell-breakdown}
\end{figure*}

\begin{table}[b!]
	\caption{Reduced $\chi^2$ results of 16 trials with various model assumptions. $\tilde{\rho}(\alpha,v)$ refers to the approximation that was used for the lightcone virtuality and momentum distribution to calculate Eq.~\ref{Eq:F2_LC}. Similarly, $f^{off}$ rerfers to the offshell functional form used. There is a systematic increase in $\chi^2$ when using $F_2^n/F_2^p$ of Arrington~\cite{Arrington:2011qt}. There is also an increase when using GCF-LC approximation to $\tilde{\rho}(\alpha,v)$. We note that when using the KP and CJ off-shell parametrization there are no free parameters fit to data and the quoted $\chi^2$ values show the quality of their description of the data without any minimization procedure.\\}
\begin{tabular}{|c|c|c|c|c|r|}
\hline
$\tilde{\rho}(\alpha,\nu)$ & $F_2^n/F_2^p$ & $f^{off}(\tilde{x})$ & $\chi^2_d$ & $\chi^2_{^3He}$ & $\chi_{tot}^2/\text{d.o.f.}$ \\ \hline
\multirow{8}{*}{SF}        & \multirow{4}{*}{Seg.}    & Const.-x             & 7.4        & 12.4            & 19.8 / 31 = 0.63             \\ \cline{3-6} 
                           &                          & Lin.-x               & 7.7        & 7.7             & 15.4 / 30 = 0.51             \\ \cline{3-6} 
                           &                          & KP                   & 12.9       & 12.1            & 25 / 32 = 0.78               \\ \cline{3-6} 
                           &                          & CJ                   & 6.6        & 23.4            & 30 / 32 = 0.94               \\ \cline{2-6} 
                           & \multirow{4}{*}{Arr.}    & Const.-x             & 17.4       & 69.1            & 86.5 / 31 = 2.79             \\ \cline{3-6} 
                           &                          & Lin.-x               & 25.9       & 16.0            & 41.9 / 30 = 1.40             \\ \cline{3-6} 
                           &                          & KP                   & 12.1       & 21.4            & 33.5 / 32 = 1.05             \\ \cline{3-6} 
                           &                          & CJ                   & 6.7        & 111.9           & 118.6 / 32 = 3.71            \\ \hline
\multirow{8}{*}{GCF-LC}    & \multirow{4}{*}{Seg.}    & Const.-x             & 8.4        & 19.2            & 27.6 / 31 = 0.89             \\ \cline{3-6} 
                           &                          & Lin.-x               & 7.2        & 16.4            & 23.6 / 30 = 0.79             \\ \cline{3-6} 
                           &                          & KP                   & 9.8        & 10.5            & 20.3 / 32 = 0.63             \\ \cline{3-6} 
                           &                          & CJ                   & 11.8       & 26.8            & 38.6 / 32 = 1.21             \\ \cline{2-6} 
                           & \multirow{4}{*}{Arr.}    & Const.-x             & 22.9       & 69.3            & 92.2 / 31 = 2.97             \\ \cline{3-6} 
                           &                          & Lin.-x               & 25.4       & 53.1            & 78.5 / 30 = 2.62             \\ \cline{3-6} 
                           &                          & KP                   & 8.7        & 64.5            & 73.2 / 32 = 2.29             \\ \cline{3-6} 
                           &                          & CJ                   & 12.9       & 110.8           & 123.7 / 32 = 3.87            \\ \hline
\end{tabular}
	\label{Tab:chi2}
\end{table}

The best fits with the Segarra et al. $F_2^n/F_2^p$ (blue curves) and Arrington (red curves) find a comparable offshell function $f^{off}$ with the LC approximation as compared to the SF approximation, see Fig.~\ref{fig:offshells}.

The GCF-LC framework does just as well at describing the $^3$He data and deuterium data, again, with improvement with the use of the Segarra et al. $F_2^n/F_2^p$. In the GCF-LC framework,
the high-$x_B$ $^3$He data not used in the fitting procedure (due to having low-$W$) is not as well described as in the SF framework.
We also note that the $f^{off}_{CJ}$ did not use $^3$He data as a constraint, and, therefore, struggles at describing the data, especially at high-$x_B$.

\subsection{SRC contribution to nucleon modification}

Using the inferred parameters from the global fit as described above, we can now separate the contributions of the mean-field and SRC nucleons to the EMC effect. To this end we constructed $F_2^A=F_2^{A}(MF)+F_2^{^A}(SRC)$ by splitting the integral in Eq.~\ref{Eq:F2_LC} to contributions of Mean-Field and SRC nucleons.
This separation is natural for the GCF-LC approach. For the SF based
approach this is done by assigning all nucleons with moment above $240$ MeV/c as members of SRC pairs. Our findings are largely insensitive to the exact momenta we choose.

Figure~\ref{fig:ratios-breakdown} shows the ratio of the structure
functions: $\Big[F_2^{^3He}(SRC)\Big]/\Big[F_2^{^3He}(MF)\Big]$ and
$\Big[F_2^{d}(MF)\Big]/\Big[F_2^{d}(SRC)\Big]$.  As expected,
mean-field nucleons account for most of the structure function in
Eq.~\ref{Eq:F2_LC}, except at very high-$x_B$ where nucleon motion
effects are important and therefore the contribution of SRCs becomes
significant.  This is to be expected as SRC nucleons account for a
small fraction of the nuclear wave function, especially in
deuterium.

Next we explicitly examine the contribution of mean-field and SRC
nucleons to the offshell modification effect in the EMC.  This is done
by defining the offshell decomposition as $F_2^A(\text{off-shell}) =
F_2^A(\text{full}) - F_2^A(\text{no off-shell})$, where
$F_2^A(\text{full})$ is calculated using Eq.~\ref{Eq:F2_LC} and
$F_2^A(\text{no off-shell})$ is calculated using the same equation but
by setting $f^{off}(\tilde{x})=0$.

Figure~\ref{fig:offshell-breakdown} shows the decomposition of
$F_2^A(\text{off-shell})$ due to SRC and mean-field nucleons within
the SF approach (LC calculations are qualitatively similar and can be
found in the online supplementary materials).  While high momentum
nucleons did not significantly contribute to the full convolution
ratio in Fig.~\ref{fig:ratios-breakdown}, these nucleons dominate the
offshell modification function (i.e., the dashed blue lines track the solid
black lines closely, especially at high $x_B$) in all models even
though the offshell behavior is different for each model.

This holds true even in deuterium at high $x_B$, although at
$x_B \sim0.6$, the mean-field and SRC contributions are closer to
1:1. This is still surprising given the high-momenta fraction of the
nuclear momentum-distribution is only $O(\sim
4\%)$~\cite{veerasamy11}. Adding to this surprise is the feature that
a significant contribution to the wave function comes from $np$ separations
larger than the range of the nuclear forces~\cite{FrankfurtStrikman:1978}.

Furthermore, in the results shown here using the SF approach, the momentum sum rule is violated by $\sim 1\%$. While small, this violation still induces an artificial EMC effect, thereby reducing the strength of the actual offshell contribution to the structure function (i.e. the absolute $y$-scale of Fig.~\ref{fig:offshell-breakdown}). Alternatively, in the LC approach, the sum rules are manifestly satisfied, and the extracted offshell contribution is much larger for the models of $F_2^n/F_2^p|_{Seg.}$ by a factor of about $1.5-3$ (see online supplementary materials).

Our findings are robust to the exact underlying offshell function used
in Eq.~\ref{Eq:F2_LC}, even though $f^{off}(\tilde{x})$
(Fig.~\ref{fig:offshells}) varies dramatically among the models.
Therefore, the results shown in Fig.~\ref{fig:offshell-breakdown}
contradict the recent claims of Ref.~\cite{Wang:2020uhj}, where the
SRC UMF was analyzed without proper separating its contributions
from nucleon motion and modification effects.  For completeness we
note that the UMF extracted by
Ref.~\cite{Schmookler:2019nvf,Segarra:2019gbp} is reproduced with the
convolution framework used here for $^3$He, see online supplementary
materials.

\section{Predicting forthcoming observables}

\begin{figure}[t]
		\includegraphics[width=0.48\textwidth,  height = 4.5cm]{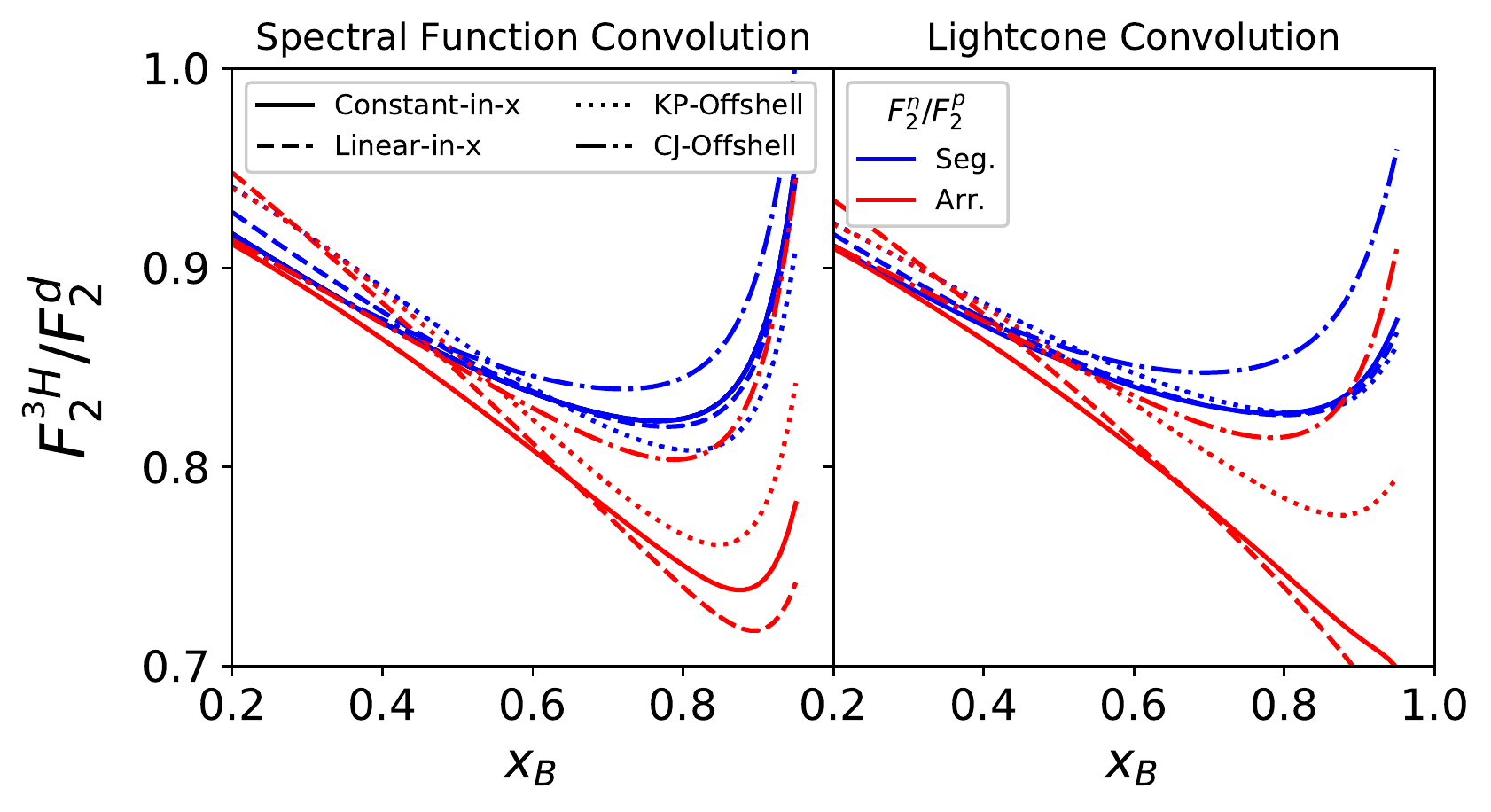}
		\caption{Predictions of $\frac{F_2^{^3H}}{F_2^d}$ using our convolution
		framework with the universal offshell modification constrained from $^2$H and $^3$He 
		data: (Left) SF convolution, (Right) LC convolution. See text for details.
		All curves are shown for MARATHON kinematics, i.e., $Q^2=14 \cdot x_B \textrm{ [GeV}^2\textrm{]}$.\\
		}
		\label{fig:h3-prediction}
\end{figure}

While existing data cannot constrain $F_2^n/F_2^p$, we predict
$F_2^{^3H}/F_2^d$, which was recently measured by the MARATHON
collaboration~\cite{MARATHON}, and should be sensitive to
$F_2^n$.  Fig.~\ref{fig:h3-prediction} shows the convolution
prediction for $F_2^{^3H}/F_2^d$ obtained using the constrained
offshell modification function and assuming isospin symmetry in the
lightcone distributions.  The different
$F_2^n/F_2^p$ parametrizations, which are both consistent with
$F_2^{^3He}/F_2^d$ data, predict very different $F_2^{^3H}/F_2^d$ at
high-$x_B$ that MARATHON can test. Still, as seen in
Fig.~\ref{fig:h3-prediction}, there are predictions of
$F_2^{^3H}/F_2^d$ which overlap for very different $F_2^n/F_2^p$ and
$f^{off}$ behaviors. In particular, taking the $f^{off}_{KP}$ with
$F_2^n/F_2^p|_{Seg.}$ (blue dotted) and $f^{off}_{CJ}$ with
$F_2^n/F_2^p|_{Arr.}$ (red dash-dotted), yield overlapping
predictions. This indicates a combined analysis of nuclear DIS data
with forthcoming data by MARATHON will be needed to disentangle
$F_2^n/F_2^p$ and $f^{off}$, similar to efforts Ref.~\cite{KULAGIN2006126}
has performed in the past.

\begin{figure}[t]
		\includegraphics[width=0.48\textwidth]{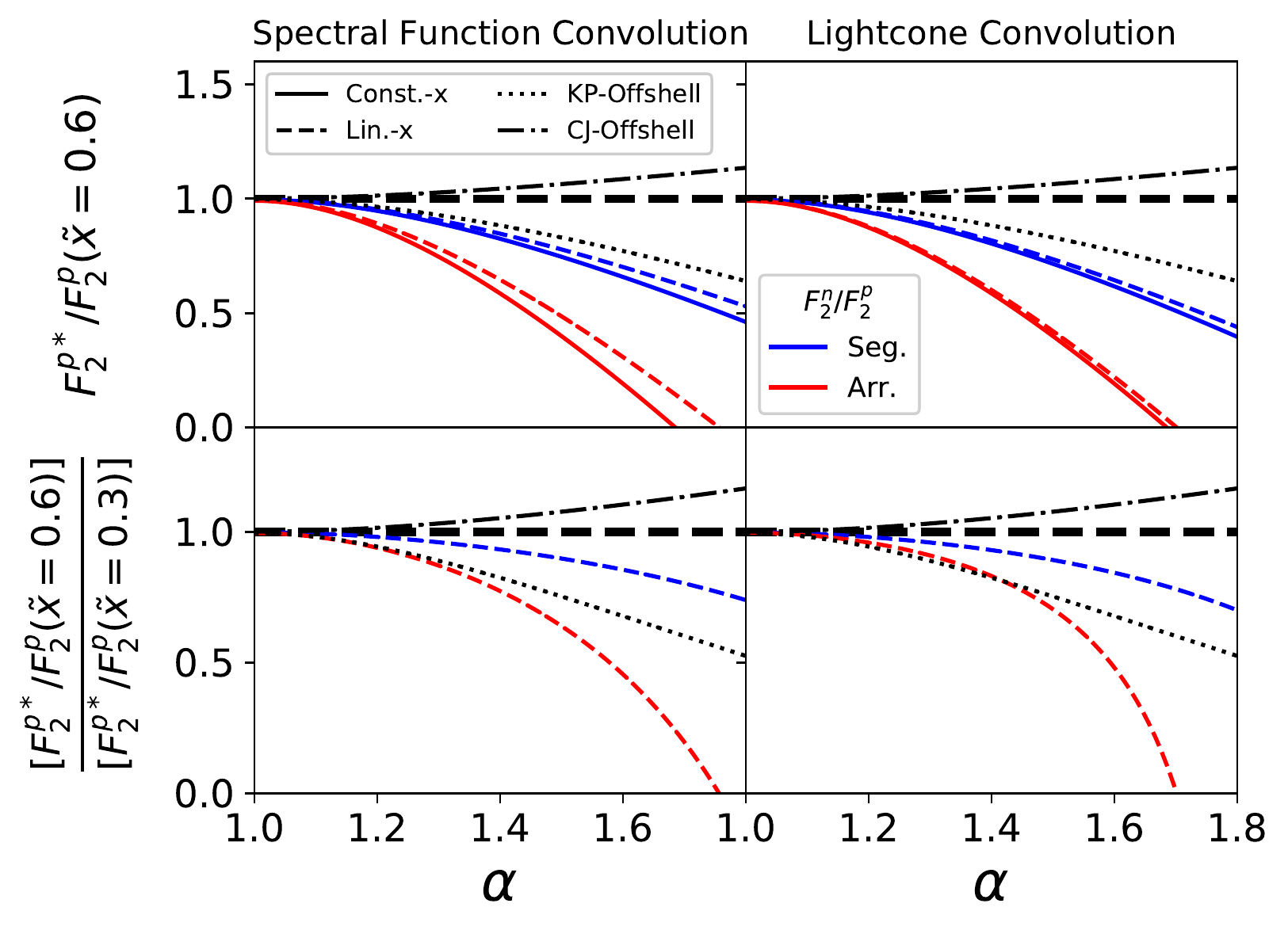}
		\caption{
		Predictions of the ratio of the bound proton structure function in deuterium 
		to the free proton structure function as a function of $\alpha$. In the bottom panels,
		predictions for models with $f^{off}_{const}$ yield a double ratio of $1$, as the modification
		is constant in $\tilde{x}$. See text for details.
		Curves are shown at $Q^2=5 \textrm{ GeV}^2$ and for ${\bf{p_T}}=0$.\\
		}
		\label{fig:boundf2}
\end{figure}

While the MARATHON results will be very sensitive to $F_2^n/F_2^p$,
they will be less  sensitive to the exact nature of the offshell modification function ($f^{off}$). 
This  can however be tested in a new set of tagged deep inelastic scattering measurements off deuterium which will study the
dependence of  the bound nucleon structure function on $\alpha$:
\begin{equation}
\begin{split}
		& F_2^{p*}(\tilde{x},\alpha)= F_2^p(\tilde{x}) \Big[ 1 + \braket{v}|_\alpha \  f^{off}(\tilde{x})\Big],
\end{split}
\end{equation}
where $\braket{v}|_\alpha$ is the average fractional virtuality for the given $\alpha$, see online supplementary materials Figs.~1 and~2.

By taking a ratio of the bound-to-free proton structure function, one can access the offshell modification function, and can examine the differences of the offshell contribution at high-$\tilde{x}$ and low-$\tilde{x}$, see Fig~\ref{fig:boundf2}. The predictions here are similar to those made by Ref.~\cite{Melnitchouk:1997pemc} and will be directly tested by the 
LAD~\cite{Emcsrcexpt11} and BAND~\cite{band-proposal} Collaborations. The latter already completed 50\% data taking and results are anticipated soon. While predictions here are made for ${\bf{p_T}}=0$, experiments will have some finite acceptance in ${\bf{p_T}}$. As seen in Fig~\ref{fig:boundf2}, there are significant uncertainties due to uncertainties in $F_2^n/F_2^p$ (red vs. blue curves). However, after precise measurements on $F_2^n/F_2^p$ by the MARATHON Collaboration~\cite{MARATHON}, these uncertainties will be greatly reduced.

\section{Summary}
We present an extensive study of nucleon modification effects in nuclei using a convolution formalism and measurements of the EMC effect in deuterium and $^3$He. We examine a range of off-shell modification functions, free-neutron structure function models and different treatments of nucleon motion effects. In all cases we find that nucleons in SRC pairs are the dominant contribution to nucleon modification effects in deuterium and $^3$He.

With upcoming precise measurements of $^{3}$H, our study can be extended to test the isospin dependence of the universal offshell modification function and the ability to use nuclear DIS data to constrain the free neutron structure function. We stress that an isospin-dependent EMC effect, in the sense of a different average modification for protons and neutrons, as e.g. suggested by Refs.~\cite{hen14,Hen:2016kwk,duer18}, can be obtained in all models discussed in this paper if the proton and neutron lightcone densities have different average virtualities. In addition we make predictions for new measurements of the bound nucleon structure function. These measurements will allow us to further constrain the elements of our model.


\begin{acknowledgments}
This work was supported by the U.S. Department of Energy, Office of Science, Office of Nuclear Physics under Award Numbers DE-FG02-94ER40818, DE-FG02-96ER-40960, DE-AC05-06OR23177, DE-FG02- 93ER-40771, and DE-FG02-97ER-41014 under which Jefferson Science Associates operates the Thomas Jefferson National Accelerator Facility, the Pazy foundation, and the Israeli Science Foundation (Israel) under Grants Nos. 136/12 and 1334/16. 
\end{acknowledgments}

\bibliography{../../../references.bib}

\begin{thebibliography}{44}%
\makeatletter
\providecommand \@ifxundefined [1]{%
 \@ifx{#1\undefined}
}%
\providecommand \@ifnum [1]{%
 \ifnum #1\expandafter \@firstoftwo
 \else \expandafter \@secondoftwo
 \fi
}%
\providecommand \@ifx [1]{%
 \ifx #1\expandafter \@firstoftwo
 \else \expandafter \@secondoftwo
 \fi
}%
\providecommand \natexlab [1]{#1}%
\providecommand \enquote  [1]{``#1''}%
\providecommand \bibnamefont  [1]{#1}%
\providecommand \bibfnamefont [1]{#1}%
\providecommand \citenamefont [1]{#1}%
\providecommand \href@noop [0]{\@secondoftwo}%
\providecommand \href [0]{\begingroup \@sanitize@url \@href}%
\providecommand \@href[1]{\@@startlink{#1}\@@href}%
\providecommand \@@href[1]{\endgroup#1\@@endlink}%
\providecommand \@sanitize@url [0]{\catcode `\\12\catcode `\$12\catcode
  `\&12\catcode `\#12\catcode `\^12\catcode `\_12\catcode `\%12\relax}%
\providecommand \@@startlink[1]{}%
\providecommand \@@endlink[0]{}%
\providecommand \url  [0]{\begingroup\@sanitize@url \@url }%
\providecommand \@url [1]{\endgroup\@href {#1}{\urlprefix }}%
\providecommand \urlprefix  [0]{URL }%
\providecommand \Eprint [0]{\href }%
\providecommand \doibase [0]{http://dx.doi.org/}%
\providecommand \selectlanguage [0]{\@gobble}%
\providecommand \bibinfo  [0]{\@secondoftwo}%
\providecommand \bibfield  [0]{\@secondoftwo}%
\providecommand \translation [1]{[#1]}%
\providecommand \BibitemOpen [0]{}%
\providecommand \bibitemStop [0]{}%
\providecommand \bibitemNoStop [0]{.\EOS\space}%
\providecommand \EOS [0]{\spacefactor3000\relax}%
\providecommand \BibitemShut  [1]{\csname bibitem#1\endcsname}%
\let\auto@bib@innerbib\@empty
\bibitem [{\citenamefont {Arnold}\ \emph {et~al.}(1984)\citenamefont {Arnold},
  \citenamefont {Bosted}, \citenamefont {Chang}, \citenamefont {Gomez},
  \citenamefont {Katramatou}, \citenamefont {Petratos}, \citenamefont {Rahbar},
  \citenamefont {Rock}, \citenamefont {Sill}, \citenamefont {Szalata},
  \citenamefont {Bodek}, \citenamefont {Giokaris}, \citenamefont {Sherden},
  \citenamefont {Mecking},\ and\ \citenamefont {Lombard}}]{Arnold:1984}%
  \BibitemOpen
  \bibfield  {author} {\bibinfo {author} {\bibfnamefont {R.~G.}\ \bibnamefont
  {Arnold}}, \bibinfo {author} {\bibfnamefont {P.~E.}\ \bibnamefont {Bosted}},
  \bibinfo {author} {\bibfnamefont {C.~C.}\ \bibnamefont {Chang}}, \bibinfo
  {author} {\bibfnamefont {J.}~\bibnamefont {Gomez}}, \bibinfo {author}
  {\bibfnamefont {A.~T.}\ \bibnamefont {Katramatou}}, \bibinfo {author}
  {\bibfnamefont {G.~G.}\ \bibnamefont {Petratos}}, \bibinfo {author}
  {\bibfnamefont {A.~A.}\ \bibnamefont {Rahbar}}, \bibinfo {author}
  {\bibfnamefont {S.~E.}\ \bibnamefont {Rock}}, \bibinfo {author}
  {\bibfnamefont {A.~F.}\ \bibnamefont {Sill}}, \bibinfo {author}
  {\bibfnamefont {Z.~M.}\ \bibnamefont {Szalata}}, \bibinfo {author}
  {\bibfnamefont {A.}~\bibnamefont {Bodek}}, \bibinfo {author} {\bibfnamefont
  {N.}~\bibnamefont {Giokaris}}, \bibinfo {author} {\bibfnamefont {D.~J.}\
  \bibnamefont {Sherden}}, \bibinfo {author} {\bibfnamefont {B.~A.}\
  \bibnamefont {Mecking}}, \ and\ \bibinfo {author} {\bibfnamefont {R.~M.}\
  \bibnamefont {Lombard}},\ }\bibfield  {title} {\enquote {\bibinfo {title}
  {Measurements of the $a$ dependence of deep-inelastic electron scattering
  from nuclei},}\ }\href {\doibase 10.1103/PhysRevLett.52.727} {\bibfield
  {journal} {\bibinfo  {journal} {Phys. Rev. Lett.}\ }\textbf {\bibinfo
  {volume} {52}},\ \bibinfo {pages} {727--730} (\bibinfo {year}
  {1984})}\BibitemShut {NoStop}%
\bibitem [{\citenamefont {Aubert}\ \emph {et~al.}(1983)\citenamefont {Aubert}
  \emph {et~al.}}]{Aubert83}%
  \BibitemOpen
  \bibfield  {author} {\bibinfo {author} {\bibfnamefont {J.J.}\ \bibnamefont
  {Aubert}} \emph {et~al.},\ }\href@noop {} {\bibfield  {journal} {\bibinfo
  {journal} {Phys. Lett. B}\ }\textbf {\bibinfo {volume} {123}},\ \bibinfo
  {pages} {275} (\bibinfo {year} {1983})}\BibitemShut {NoStop}%
\bibitem [{\citenamefont {Ashman}\ \emph {et~al.}(1988)\citenamefont {Ashman}
  \emph {et~al.}}]{Ashman88}%
  \BibitemOpen
  \bibfield  {author} {\bibinfo {author} {\bibfnamefont {J.}~\bibnamefont
  {Ashman}} \emph {et~al.},\ }\bibfield  {title} {\enquote {\bibinfo {title}
  {Measurement of the ratios of deep inelastic muon-nucleus cross sections on
  various nuclei compared to deuterium},}\ }\href@noop {} {\bibfield  {journal}
  {\bibinfo  {journal} {Phys. Lett. B}\ }\textbf {\bibinfo {volume} {202}},\
  \bibinfo {pages} {603} (\bibinfo {year} {1988})}\BibitemShut {NoStop}%
\bibitem [{\citenamefont {Gomez}\ \emph {et~al.}(1994)\citenamefont {Gomez}
  \emph {et~al.}}]{Gomez94}%
  \BibitemOpen
  \bibfield  {author} {\bibinfo {author} {\bibfnamefont {J.}~\bibnamefont
  {Gomez}} \emph {et~al.},\ }\bibfield  {title} {\enquote {\bibinfo {title}
  {Measurement of the $a$ dependence of deep-inelastic electron scattering},}\
  }\href@noop {} {\bibfield  {journal} {\bibinfo  {journal} {Phys. Rev. D}\
  }\textbf {\bibinfo {volume} {49}},\ \bibinfo {pages} {4348} (\bibinfo {year}
  {1994})}\BibitemShut {NoStop}%
\bibitem [{\citenamefont {Arneodo}\ \emph {et~al.}(1990)\citenamefont {Arneodo}
  \emph {et~al.}}]{Arneodo90}%
  \BibitemOpen
  \bibfield  {author} {\bibinfo {author} {\bibfnamefont {M.}~\bibnamefont
  {Arneodo}} \emph {et~al.},\ }\bibfield  {title} {\enquote {\bibinfo {title}
  {Measurements of the nucleon structure function in the range $0.002 < x <
  0.17$ and $0.2 < q^2 < 8$ gev$^2$ in deuterium, carbon and calcium},}\
  }\href@noop {} {\bibfield  {journal} {\bibinfo  {journal} {Nucl. Phys. B}\
  }\textbf {\bibinfo {volume} {333}},\ \bibinfo {pages} {1} (\bibinfo {year}
  {1990})}\BibitemShut {NoStop}%
\bibitem [{\citenamefont {Seely}\ \emph {et~al.}(2009)\citenamefont {Seely}
  \emph {et~al.}}]{Seely09}%
  \BibitemOpen
  \bibfield  {author} {\bibinfo {author} {\bibfnamefont {J.}~\bibnamefont
  {Seely}} \emph {et~al.},\ }\bibfield  {title} {\enquote {\bibinfo {title}
  {New measurements of the european muon collaboration effect in very light
  nuclei},}\ }\href@noop {} {\bibfield  {journal} {\bibinfo  {journal} {Phys.
  Rev. Lett.}\ }\textbf {\bibinfo {volume} {103}},\ \bibinfo {pages} {202301}
  (\bibinfo {year} {2009})}\BibitemShut {NoStop}%
\bibitem [{\citenamefont {Schmookler}\ \emph {et~al.}(2019)\citenamefont
  {Schmookler} \emph {et~al.}}]{Schmookler:2019nvf}%
  \BibitemOpen
  \bibfield  {author} {\bibinfo {author} {\bibfnamefont {B.}~\bibnamefont
  {Schmookler}} \emph {et~al.} (\bibinfo {collaboration} {CLAS
  Collaboration}),\ }\bibfield  {title} {\enquote {\bibinfo {title} {{Modified
  structure of protons and neutrons in correlated pairs}},}\ }\href {\doibase
  10.1038/s41586-019-0925-9} {\bibfield  {journal} {\bibinfo  {journal}
  {Nature}\ }\textbf {\bibinfo {volume} {566}},\ \bibinfo {pages} {354--358}
  (\bibinfo {year} {2019})}\BibitemShut {NoStop}%
\bibitem [{\citenamefont {Frankfurt}\ and\ \citenamefont
  {Strikman}(1988)}]{Frankfurt88}%
  \BibitemOpen
  \bibfield  {author} {\bibinfo {author} {\bibfnamefont {Leonid}\ \bibnamefont
  {Frankfurt}}\ and\ \bibinfo {author} {\bibfnamefont {Mark}\ \bibnamefont
  {Strikman}},\ }\bibfield  {title} {\enquote {\bibinfo {title} {Hard nuclear
  processes and microscopic nuclear structure},}\ }\href@noop {} {\bibfield
  {journal} {\bibinfo  {journal} {Phys. Rep.}\ }\textbf {\bibinfo {volume}
  {160}},\ \bibinfo {pages} {235 -- 427} (\bibinfo {year} {1988})}\BibitemShut
  {NoStop}%
\bibitem [{\citenamefont {Norton}(2003)}]{Norton03}%
  \BibitemOpen
  \bibfield  {author} {\bibinfo {author} {\bibfnamefont {P~R}\ \bibnamefont
  {Norton}},\ }\bibfield  {title} {\enquote {\bibinfo {title} {The {EMC}
  effect},}\ }\href {\doibase 10.1088/0034-4885/66/8/201} {\bibfield  {journal}
  {\bibinfo  {journal} {Reports on Progress in Physics}\ }\textbf {\bibinfo
  {volume} {66}},\ \bibinfo {pages} {1253--1297} (\bibinfo {year}
  {2003})}\BibitemShut {NoStop}%
\bibitem [{\citenamefont {Hen}\ \emph {et~al.}(2017)\citenamefont {Hen},
  \citenamefont {Miller}, \citenamefont {Piasetzky},\ and\ \citenamefont
  {Weinstein}}]{Hen:2016kwk}%
  \BibitemOpen
  \bibfield  {author} {\bibinfo {author} {\bibfnamefont {O.}~\bibnamefont
  {Hen}}, \bibinfo {author} {\bibfnamefont {G.~A.}\ \bibnamefont {Miller}},
  \bibinfo {author} {\bibfnamefont {E.}~\bibnamefont {Piasetzky}}, \ and\
  \bibinfo {author} {\bibfnamefont {L.~B.}\ \bibnamefont {Weinstein}},\
  }\bibfield  {title} {\enquote {\bibinfo {title} {{Nucleon-Nucleon
  Correlations, Short-lived Excitations, and the Quarks Within}},}\ }\href
  {\doibase 10.1103/RevModPhys.89.045002} {\bibfield  {journal} {\bibinfo
  {journal} {Rev. Mod. Phys.}\ }\textbf {\bibinfo {volume} {89}},\ \bibinfo
  {pages} {045002} (\bibinfo {year} {2017})}\BibitemShut {NoStop}%
\bibitem [{\citenamefont {Alde}\ \emph {et~al.}(1990)\citenamefont {Alde} \emph
  {et~al.}}]{Alde:1990im}%
  \BibitemOpen
  \bibfield  {author} {\bibinfo {author} {\bibfnamefont {D.~M.}\ \bibnamefont
  {Alde}} \emph {et~al.},\ }\bibfield  {title} {\enquote {\bibinfo {title}
  {{Nuclear dependence of dimuon production at 800-GeV. FNAL-772
  experiment}},}\ }\href {\doibase 10.1103/PhysRevLett.64.2479} {\bibfield
  {journal} {\bibinfo  {journal} {Phys. Rev. Lett.}\ }\textbf {\bibinfo
  {volume} {64}},\ \bibinfo {pages} {2479--2482} (\bibinfo {year}
  {1990})}\BibitemShut {NoStop}%
\bibitem [{\citenamefont {Weinstein}\ \emph {et~al.}(2011)\citenamefont
  {Weinstein}, \citenamefont {Piasetzky}, \citenamefont {Higinbotham},
  \citenamefont {Gomez}, \citenamefont {Hen},\ and\ \citenamefont
  {Shneor}}]{weinstein11}%
  \BibitemOpen
  \bibfield  {author} {\bibinfo {author} {\bibfnamefont {L.~B.}\ \bibnamefont
  {Weinstein}}, \bibinfo {author} {\bibfnamefont {E.}~\bibnamefont
  {Piasetzky}}, \bibinfo {author} {\bibfnamefont {D.~W.}\ \bibnamefont
  {Higinbotham}}, \bibinfo {author} {\bibfnamefont {J.}~\bibnamefont {Gomez}},
  \bibinfo {author} {\bibfnamefont {O.}~\bibnamefont {Hen}}, \ and\ \bibinfo
  {author} {\bibfnamefont {R.}~\bibnamefont {Shneor}},\ }\bibfield  {title}
  {\enquote {\bibinfo {title} {Short range correlations and the emc effect},}\
  }\href {\doibase 10.1103/PhysRevLett.106.052301} {\bibfield  {journal}
  {\bibinfo  {journal} {Phys. Rev. Lett.}\ }\textbf {\bibinfo {volume} {106}},\
  \bibinfo {pages} {052301} (\bibinfo {year} {2011})}\BibitemShut {NoStop}%
\bibitem [{\citenamefont {Hen}\ \emph {et~al.}(2012)\citenamefont {Hen},
  \citenamefont {Piasetzky},\ and\ \citenamefont {Weinstein}}]{Hen12}%
  \BibitemOpen
  \bibfield  {author} {\bibinfo {author} {\bibfnamefont {O.}~\bibnamefont
  {Hen}}, \bibinfo {author} {\bibfnamefont {E.}~\bibnamefont {Piasetzky}}, \
  and\ \bibinfo {author} {\bibfnamefont {L.~B.}\ \bibnamefont {Weinstein}},\
  }\bibfield  {title} {\enquote {\bibinfo {title} {New data strengthen the
  connection between short range correlations and the emc effect},}\ }\href
  {\doibase 10.1103/PhysRevC.85.047301} {\bibfield  {journal} {\bibinfo
  {journal} {Phys. Rev. C}\ }\textbf {\bibinfo {volume} {85}},\ \bibinfo
  {pages} {047301} (\bibinfo {year} {2012})}\BibitemShut {NoStop}%
\bibitem [{\citenamefont {Hen}\ \emph {et~al.}(2013)\citenamefont {Hen},
  \citenamefont {Higinbotham}, \citenamefont {Miller}, \citenamefont
  {Piasetzky},\ and\ \citenamefont {Weinstein}}]{Hen:2013oha}%
  \BibitemOpen
  \bibfield  {author} {\bibinfo {author} {\bibfnamefont {Or}~\bibnamefont
  {Hen}}, \bibinfo {author} {\bibfnamefont {D.~W.}\ \bibnamefont
  {Higinbotham}}, \bibinfo {author} {\bibfnamefont {Gerald~A.}\ \bibnamefont
  {Miller}}, \bibinfo {author} {\bibfnamefont {Eli}\ \bibnamefont {Piasetzky}},
  \ and\ \bibinfo {author} {\bibfnamefont {Lawrence~B.}\ \bibnamefont
  {Weinstein}},\ }\bibfield  {title} {\enquote {\bibinfo {title} {{The EMC
  Effect and High Momentum Nucleons in Nuclei}},}\ }\href {\doibase
  10.1142/S0218301313300178} {\bibfield  {journal} {\bibinfo  {journal} {Int.
  J. Mod. Phys.}\ }\textbf {\bibinfo {volume} {E22}},\ \bibinfo {pages}
  {1330017} (\bibinfo {year} {2013})},\ \Eprint
  {http://arxiv.org/abs/1304.2813} {arXiv:1304.2813 [nucl-th]} \BibitemShut
  {NoStop}%
\bibitem [{\citenamefont {Segarra}\ \emph {et~al.}(2020)\citenamefont
  {Segarra}, \citenamefont {Schmidt}, \citenamefont {Higinbotham},
  \citenamefont {Piasetzky}, \citenamefont {Strikman}, \citenamefont
  {Weinstein},\ and\ \citenamefont {Hen}}]{Segarra:2019gbp}%
  \BibitemOpen
  \bibfield  {author} {\bibinfo {author} {\bibfnamefont {E.~P.}\ \bibnamefont
  {Segarra}}, \bibinfo {author} {\bibfnamefont {A.}~\bibnamefont {Schmidt}},
  \bibinfo {author} {\bibfnamefont {D.~W.}\ \bibnamefont {Higinbotham}},
  \bibinfo {author} {\bibfnamefont {E.}~\bibnamefont {Piasetzky}}, \bibinfo
  {author} {\bibfnamefont {M.}~\bibnamefont {Strikman}}, \bibinfo {author}
  {\bibfnamefont {L.~B.}\ \bibnamefont {Weinstein}}, \ and\ \bibinfo {author}
  {\bibfnamefont {O.}~\bibnamefont {Hen}},\ }\bibfield  {title} {\enquote
  {\bibinfo {title} {{Flavor dependence of the nucleon valence structure from
  nuclear deep inelastic scattering data}},}\ }\href@noop {} {\bibfield
  {journal} {\bibinfo  {journal} {Phys. Rev. Lett.}\ } (\bibinfo {year}
  {2020})},\ \Eprint {http://arxiv.org/abs/1908.02223} {arXiv:1908.02223
  [nucl-th]} \BibitemShut {NoStop}%
\bibitem [{\citenamefont {Petratos}\ \emph {et~al.}(2010)\citenamefont
  {Petratos} \emph {et~al.}}]{MARATHON}%
  \BibitemOpen
  \bibfield  {author} {\bibinfo {author} {\bibfnamefont {G.~G.}\ \bibnamefont
  {Petratos}} \emph {et~al.},\ }\bibfield  {title} {\enquote {\bibinfo {title}
  {{MeAsurement of the $F_2^n/F_2^p$, $d/u$ RAtios and $A=3$ EMC Effect in Deep
  Inelastic Electron Scattering Off the Tritium and Helium MirrOr Nuclei}},}\
  }\href@noop {} {\bibfield  {journal} {\bibinfo  {journal} {Jefferson Lab
  PAC36 Proposal}\ } (\bibinfo {year} {2010})}\BibitemShut {NoStop}%
\bibitem [{\citenamefont {Hen}\ \emph {et~al.}(2015)\citenamefont {Hen} \emph
  {et~al.}}]{band-proposal}%
  \BibitemOpen
  \bibfield  {author} {\bibinfo {author} {\bibfnamefont {O.}~\bibnamefont
  {Hen}} \emph {et~al.},\ }\href
  {https://misportal.jlab.org/pacProposals/proposals/1159/attachments/93263/E12-11-003A.pdf}
  {\enquote {\bibinfo {title} {In medium proton structure functions, src, and
  the emc effect},}\ }\bibinfo {howpublished} {JLab PAC Proposal} (\bibinfo
  {year} {2015})\BibitemShut {NoStop}%
\bibitem [{\citenamefont {Hen}\ \emph {et~al.}(2011)\citenamefont {Hen},
  \citenamefont {Weinstein}, \citenamefont {Wood},\ and\ \citenamefont
  {Gilad}}]{Emcsrcexpt11}%
  \BibitemOpen
  \bibfield  {author} {\bibinfo {author} {\bibfnamefont {O.}~\bibnamefont
  {Hen}}, \bibinfo {author} {\bibfnamefont {L.B.}\ \bibnamefont {Weinstein}},
  \bibinfo {author} {\bibfnamefont {S.A.}\ \bibnamefont {Wood}}, \ and\
  \bibinfo {author} {\bibfnamefont {S.}~\bibnamefont {Gilad}},\ }\href@noop {}
  {\enquote {\bibinfo {title} {{In Medium Nucleon Structure Functions, SRC, and
  the EMC effect, Jefferson Lab experiment E12-11-107}},}\ } (\bibinfo {year}
  {2011})\BibitemShut {NoStop}%
\bibitem [{\citenamefont {Frankfurt}\ and\ \citenamefont
  {Strikman}(1985)}]{Frankfurt:1985cv}%
  \BibitemOpen
  \bibfield  {author} {\bibinfo {author} {\bibfnamefont {L.~L.}\ \bibnamefont
  {Frankfurt}}\ and\ \bibinfo {author} {\bibfnamefont {M.~I.}\ \bibnamefont
  {Strikman}},\ }\bibfield  {title} {\enquote {\bibinfo {title} {{Point-like
  Configurations in Hadrons and Nuclei and Deep Inelastic Reactions with
  Leptons: EMC and EMC Like Effects}},}\ }\href {\doibase
  10.1016/0550-3213(85)90477-8} {\bibfield  {journal} {\bibinfo  {journal}
  {Nucl. Phys.}\ }\textbf {\bibinfo {volume} {B250}},\ \bibinfo {pages}
  {143--176} (\bibinfo {year} {1985})}\BibitemShut {NoStop}%
\bibitem [{\citenamefont {Frankfurt}\ and\ \citenamefont
  {Strikman}(1987)}]{FSemc:1987}%
  \BibitemOpen
  \bibfield  {author} {\bibinfo {author} {\bibfnamefont {L.L.}\ \bibnamefont
  {Frankfurt}}\ and\ \bibinfo {author} {\bibfnamefont {M.I.}\ \bibnamefont
  {Strikman}},\ }\bibfield  {title} {\enquote {\bibinfo {title} {{On the
  normalization of nucleus spectral function and the EMC effect}},}\ }\href
  {\doibase 10.1016/0370-2693(87)90958-0} {\bibfield  {journal} {\bibinfo
  {journal} {Phys. Lett.}\ }\textbf {\bibinfo {volume} {B183}},\ \bibinfo
  {pages} {254} (\bibinfo {year} {1987})}\BibitemShut {NoStop}%
\bibitem [{\citenamefont {Akulinichev}\ \emph {et~al.}(1985)\citenamefont
  {Akulinichev}, \citenamefont {Shlomo}, \citenamefont {Kulagin},\ and\
  \citenamefont {Vagradov}}]{SAKulaginSVAku:1985}%
  \BibitemOpen
  \bibfield  {author} {\bibinfo {author} {\bibfnamefont {S.V.}\ \bibnamefont
  {Akulinichev}}, \bibinfo {author} {\bibfnamefont {S.}~\bibnamefont {Shlomo}},
  \bibinfo {author} {\bibfnamefont {S.A.}\ \bibnamefont {Kulagin}}, \ and\
  \bibinfo {author} {\bibfnamefont {G.M.}\ \bibnamefont {Vagradov}},\
  }\bibfield  {title} {\enquote {\bibinfo {title} {{Lepton-Nucleus
  Deep-Inelastic Scattering}},}\ }\href {\doibase 10.1103/PhysRevLett.55.2239}
  {\bibfield  {journal} {\bibinfo  {journal} {Phys. Rev. Lett.}\ }\textbf
  {\bibinfo {volume} {55}},\ \bibinfo {pages} {2239} (\bibinfo {year}
  {1985})}\BibitemShut {NoStop}%
\bibitem [{\citenamefont {Dunne}\ and\ \citenamefont
  {Thomas}(1985)}]{DunneThomas:1985}%
  \BibitemOpen
  \bibfield  {author} {\bibinfo {author} {\bibfnamefont {G.V.}\ \bibnamefont
  {Dunne}}\ and\ \bibinfo {author} {\bibfnamefont {A.W.}\ \bibnamefont
  {Thomas}},\ }\bibfield  {title} {\enquote {\bibinfo {title} {{The Effect of
  Conventional Nuclear Binding on Nuclear Structure Functions}},}\ }\href
  {\doibase 10.1016/0375-9474(86)90458-6} {\bibfield  {journal} {\bibinfo
  {journal} {Nucl.Phys.}\ }\textbf {\bibinfo {volume} {A455}},\ \bibinfo
  {pages} {701--719} (\bibinfo {year} {1985})}\BibitemShut {NoStop}%
\bibitem [{\citenamefont {Miller}(2019)}]{Miller:2019mae}%
  \BibitemOpen
  \bibfield  {author} {\bibinfo {author} {\bibfnamefont {Gerald~A.}\
  \bibnamefont {Miller}},\ }\bibfield  {title} {\enquote {\bibinfo {title}
  {{Confinement in Nuclei and the Expanding Proton}},}\ }\href {\doibase
  10.1103/PhysRevLett.123.232003} {\bibfield  {journal} {\bibinfo  {journal}
  {Phys. Rev. Lett.}\ }\textbf {\bibinfo {volume} {123}},\ \bibinfo {pages}
  {232003} (\bibinfo {year} {2019})},\ \Eprint
  {http://arxiv.org/abs/1907.00110} {arXiv:1907.00110 [nucl-th]} \BibitemShut
  {NoStop}%
\bibitem [{\citenamefont {Jung}\ and\ \citenamefont
  {Miller}(1990)}]{Miller:1990}%
  \BibitemOpen
  \bibfield  {author} {\bibinfo {author} {\bibfnamefont {H.}~\bibnamefont
  {Jung}}\ and\ \bibinfo {author} {\bibfnamefont {Gerald~A.}\ \bibnamefont
  {Miller}},\ }\bibfield  {title} {\enquote {\bibinfo {title} {{Pionic
  contributions to deep inelastic nuclear structure functions}},}\ }\href
  {\doibase 10.1103/PhysRevC.41.659} {\bibfield  {journal} {\bibinfo  {journal}
  {Phys. Rev.}\ }\textbf {\bibinfo {volume} {C41}},\ \bibinfo {pages} {659}
  (\bibinfo {year} {1990})}\BibitemShut {NoStop}%
\bibitem [{\citenamefont {Jung}\ and\ \citenamefont
  {Miller}(1988)}]{Miller:1988}%
  \BibitemOpen
  \bibfield  {author} {\bibinfo {author} {\bibfnamefont {H.}~\bibnamefont
  {Jung}}\ and\ \bibinfo {author} {\bibfnamefont {Gerald~A.}\ \bibnamefont
  {Miller}},\ }\bibfield  {title} {\enquote {\bibinfo {title} {{Nucleonic
  contribution to lepton-nucleus deep inelastic scattering}},}\ }\href
  {\doibase 10.1016/0370-2693(88)90786-1} {\bibfield  {journal} {\bibinfo
  {journal} {Phys. Lett.}\ }\textbf {\bibinfo {volume} {B200}},\ \bibinfo
  {pages} {351} (\bibinfo {year} {1988})}\BibitemShut {NoStop}%
\bibitem [{\citenamefont {Ciofi~degli Atti}\ and\ \citenamefont
  {Kaptari}(2005)}]{AttiKaptari:2005}%
  \BibitemOpen
  \bibfield  {author} {\bibinfo {author} {\bibfnamefont {C.}~\bibnamefont
  {Ciofi~degli Atti}}\ and\ \bibinfo {author} {\bibfnamefont {L.P.}\
  \bibnamefont {Kaptari}},\ }\bibfield  {title} {\enquote {\bibinfo {title}
  {{Calculations of the Exclusive Processes 2H(e,e'p)n, 3He(e,e'p)2H and
  3He(e,e'p)(pn) within a Generalized Glauber Approach}},}\ }\href {\doibase
  10.1103/PhysRevC.71.024005} {\bibfield  {journal} {\bibinfo  {journal} {Phys.
  Rev.}\ }\textbf {\bibinfo {volume} {C71}} (\bibinfo {year} {2005}),\
  10.1103/PhysRevC.71.024005}\BibitemShut {NoStop}%
\bibitem [{\citenamefont {Ciofi~degli Atti}\ and\ \citenamefont
  {Simula}(1996)}]{CiofidegliAtti:1995qe}%
  \BibitemOpen
  \bibfield  {author} {\bibinfo {author} {\bibfnamefont {Claudio}\ \bibnamefont
  {Ciofi~degli Atti}}\ and\ \bibinfo {author} {\bibfnamefont {S.}~\bibnamefont
  {Simula}},\ }\bibfield  {title} {\enquote {\bibinfo {title} {{Realistic model
  of the nucleon spectral function in few and many nucleon systems}},}\ }\href
  {\doibase 10.1103/PhysRevC.53.1689} {\bibfield  {journal} {\bibinfo
  {journal} {Phys. Rev. C}\ }\textbf {\bibinfo {volume} {53}},\ \bibinfo
  {pages} {1689} (\bibinfo {year} {1996})}\BibitemShut {NoStop}%
\bibitem [{\citenamefont {{C. Ciofi degli Atti, L.L. Frankfurt, L.P. Kaptari
  and M.I. Strikman}}(2007)}]{Ciofi07}%
  \BibitemOpen
  \bibfield  {author} {\bibinfo {author} {\bibnamefont {{C. Ciofi degli Atti,
  L.L. Frankfurt, L.P. Kaptari and M.I. Strikman}}},\ }\bibfield  {title}
  {\enquote {\bibinfo {title} {On the dependence of the wave function of a
  bound nucleon on its momentum and the emc effect},}\ }\href@noop {}
  {\bibfield  {journal} {\bibinfo  {journal} {Phys. Rev. C}\ }\textbf {\bibinfo
  {volume} {76}},\ \bibinfo {pages} {055206} (\bibinfo {year}
  {2007})}\BibitemShut {NoStop}%
\bibitem [{\citenamefont {Weiss}\ \emph {et~al.}(2019)\citenamefont {Weiss},
  \citenamefont {Korover}, \citenamefont {Piasetzky}, \citenamefont {Hen},\
  and\ \citenamefont {Barnea}}]{Weiss:2018tbu}%
  \BibitemOpen
  \bibfield  {author} {\bibinfo {author} {\bibfnamefont {Ronen}\ \bibnamefont
  {Weiss}}, \bibinfo {author} {\bibfnamefont {Igor}\ \bibnamefont {Korover}},
  \bibinfo {author} {\bibfnamefont {Eliezer}\ \bibnamefont {Piasetzky}},
  \bibinfo {author} {\bibfnamefont {Or}~\bibnamefont {Hen}}, \ and\ \bibinfo
  {author} {\bibfnamefont {Nir}\ \bibnamefont {Barnea}},\ }\bibfield  {title}
  {\enquote {\bibinfo {title} {{Energy and momentum dependence of nuclear
  short-range correlations - Spectral function, exclusive scattering
  experiments and the contact formalism}},}\ }\href {\doibase
  10.1016/j.physletb.2019.02.019} {\bibfield  {journal} {\bibinfo  {journal}
  {Phys. Lett.}\ }\textbf {\bibinfo {volume} {B791}},\ \bibinfo {pages}
  {242--248} (\bibinfo {year} {2019})},\ \Eprint
  {http://arxiv.org/abs/1806.10217} {arXiv:1806.10217 [nucl-th]} \BibitemShut
  {NoStop}%
\bibitem [{\citenamefont {Pybus}\ \emph {et~al.}(2020)\citenamefont {Pybus},
  \citenamefont {Korover}, \citenamefont {Weiss}, \citenamefont {Schmidt},
  \citenamefont {Barnea}, \citenamefont {Higinbotham}, \citenamefont
  {Piasetzky}, \citenamefont {Strikman}, \citenamefont {Weinstein},\ and\
  \citenamefont {Hen}}]{Pybus:2020itv}%
  \BibitemOpen
  \bibfield  {author} {\bibinfo {author} {\bibfnamefont {J.~R.}\ \bibnamefont
  {Pybus}}, \bibinfo {author} {\bibfnamefont {I.}~\bibnamefont {Korover}},
  \bibinfo {author} {\bibfnamefont {R.}~\bibnamefont {Weiss}}, \bibinfo
  {author} {\bibfnamefont {A.}~\bibnamefont {Schmidt}}, \bibinfo {author}
  {\bibfnamefont {N.}~\bibnamefont {Barnea}}, \bibinfo {author} {\bibfnamefont
  {D.~W.}\ \bibnamefont {Higinbotham}}, \bibinfo {author} {\bibfnamefont
  {E.}~\bibnamefont {Piasetzky}}, \bibinfo {author} {\bibfnamefont
  {M.}~\bibnamefont {Strikman}}, \bibinfo {author} {\bibfnamefont {L.~B.}\
  \bibnamefont {Weinstein}}, \ and\ \bibinfo {author} {\bibfnamefont
  {O.}~\bibnamefont {Hen}},\ }\bibfield  {title} {\enquote {\bibinfo {title}
  {{Generalized Contact Formalism Analysis of the $^4$He$(e,e'pN)$
  Reaction}},}\ }\href@noop {} {\  (\bibinfo {year} {2020})},\ \Eprint
  {http://arxiv.org/abs/2003.02318} {arXiv:2003.02318 [nucl-th]} \BibitemShut
  {NoStop}%
\bibitem [{\citenamefont {Schmidt}\ \emph {et~al.}(2020)\citenamefont {Schmidt}
  \emph {et~al.}}]{schmidt20}%
  \BibitemOpen
  \bibfield  {author} {\bibinfo {author} {\bibfnamefont {A.}~\bibnamefont
  {Schmidt}} \emph {et~al.} (\bibinfo {collaboration} {CLAS}),\ }\bibfield
  {title} {\enquote {\bibinfo {title} {{Probing the core of the strong nuclear
  interaction}},}\ }\href {\doibase 10.1038/s41586-020-2021-6} {\bibfield
  {journal} {\bibinfo  {journal} {Nature}\ }\textbf {\bibinfo {volume} {578}},\
  \bibinfo {pages} {540--544} (\bibinfo {year} {2020})},\ \Eprint
  {http://arxiv.org/abs/2004.11221} {arXiv:2004.11221 [nucl-ex]} \BibitemShut
  {NoStop}%
\bibitem [{\citenamefont {Weiss}\ \emph {et~al.}(2018)\citenamefont {Weiss},
  \citenamefont {Cruz-Torres}, \citenamefont {Barnea}, \citenamefont
  {Piasetzky},\ and\ \citenamefont {Hen}}]{Weiss:2016obx}%
  \BibitemOpen
  \bibfield  {author} {\bibinfo {author} {\bibfnamefont {R.}~\bibnamefont
  {Weiss}}, \bibinfo {author} {\bibfnamefont {R.}~\bibnamefont {Cruz-Torres}},
  \bibinfo {author} {\bibfnamefont {N.}~\bibnamefont {Barnea}}, \bibinfo
  {author} {\bibfnamefont {E.}~\bibnamefont {Piasetzky}}, \ and\ \bibinfo
  {author} {\bibfnamefont {O.}~\bibnamefont {Hen}},\ }\bibfield  {title}
  {\enquote {\bibinfo {title} {{The nuclear contacts and short range
  correlations in nuclei}},}\ }\href@noop {} {\bibfield  {journal} {\bibinfo
  {journal} {Phys. Lett. B}\ }\textbf {\bibinfo {volume} {780}},\ \bibinfo
  {pages} {211} (\bibinfo {year} {2018})}\BibitemShut {NoStop}%
\bibitem [{\citenamefont {Cruz-Torres}\ \emph {et~al.}(2019)\citenamefont
  {Cruz-Torres}, \citenamefont {Lonardoni}, \citenamefont {Weiss},
  \citenamefont {Barnea}, \citenamefont {Higinbotham}, \citenamefont
  {Piasetzky}, \citenamefont {Schmidt}, \citenamefont {Weinstein},
  \citenamefont {Wiringa},\ and\ \citenamefont {Hen}}]{Cruz-Torres:2019fum}%
  \BibitemOpen
  \bibfield  {author} {\bibinfo {author} {\bibfnamefont {R.}~\bibnamefont
  {Cruz-Torres}}, \bibinfo {author} {\bibfnamefont {D.}~\bibnamefont
  {Lonardoni}}, \bibinfo {author} {\bibfnamefont {R.}~\bibnamefont {Weiss}},
  \bibinfo {author} {\bibfnamefont {N.}~\bibnamefont {Barnea}}, \bibinfo
  {author} {\bibfnamefont {D.~W.}\ \bibnamefont {Higinbotham}}, \bibinfo
  {author} {\bibfnamefont {E.}~\bibnamefont {Piasetzky}}, \bibinfo {author}
  {\bibfnamefont {A.}~\bibnamefont {Schmidt}}, \bibinfo {author} {\bibfnamefont
  {L.~B.}\ \bibnamefont {Weinstein}}, \bibinfo {author} {\bibfnamefont {R.~B.}\
  \bibnamefont {Wiringa}}, \ and\ \bibinfo {author} {\bibfnamefont
  {O.}~\bibnamefont {Hen}},\ }\bibfield  {title} {\enquote {\bibinfo {title}
  {{Scale and Scheme Independence and Position-Momentum Equivalence of Nuclear
  Short-Range Correlations}},}\ }\href@noop {} {\bibfield  {journal} {\bibinfo
  {journal} {arXiv}\ } (\bibinfo {year} {2019})},\ \Eprint
  {http://arxiv.org/abs/1907.03658} {arXiv:1907.03658 [nucl-th]} \BibitemShut
  {NoStop}%
\bibitem [{\citenamefont {Kulagin}\ and\ \citenamefont
  {Petti}(2006)}]{KULAGIN2006126}%
  \BibitemOpen
  \bibfield  {author} {\bibinfo {author} {\bibfnamefont {S.A.}\ \bibnamefont
  {Kulagin}}\ and\ \bibinfo {author} {\bibfnamefont {R.}~\bibnamefont
  {Petti}},\ }\bibfield  {title} {\enquote {\bibinfo {title} {Global study of
  nuclear structure functions},}\ }\href {\doibase
  http://dx.doi.org/10.1016/j.nuclphysa.2005.10.011} {\bibfield  {journal}
  {\bibinfo  {journal} {Nuclear Physics A}\ }\textbf {\bibinfo {volume}
  {765}},\ \bibinfo {pages} {126 -- 187} (\bibinfo {year} {2006})}\BibitemShut
  {NoStop}%
\bibitem [{\citenamefont {Accardi}\ \emph {et~al.}(2016)\citenamefont
  {Accardi}, \citenamefont {Brady}, \citenamefont {Melnitchouk}, \citenamefont
  {Owens},\ and\ \citenamefont {Sato}}]{Accardi:2016qay}%
  \BibitemOpen
  \bibfield  {author} {\bibinfo {author} {\bibfnamefont {A.}~\bibnamefont
  {Accardi}}, \bibinfo {author} {\bibfnamefont {L.~T.}\ \bibnamefont {Brady}},
  \bibinfo {author} {\bibfnamefont {W.}~\bibnamefont {Melnitchouk}}, \bibinfo
  {author} {\bibfnamefont {J.~F.}\ \bibnamefont {Owens}}, \ and\ \bibinfo
  {author} {\bibfnamefont {N.}~\bibnamefont {Sato}},\ }\bibfield  {title}
  {\enquote {\bibinfo {title} {{Constraints on large-$x$ parton distributions
  from new weak boson production and deep-inelastic scattering data}},}\ }\href
  {\doibase 10.1103/PhysRevD.93.114017} {\bibfield  {journal} {\bibinfo
  {journal} {Phys. Rev.}\ }\textbf {\bibinfo {volume} {D93}},\ \bibinfo {pages}
  {114017} (\bibinfo {year} {2016})},\ \Eprint
  {http://arxiv.org/abs/1602.03154} {arXiv:1602.03154 [hep-ph]} \BibitemShut
  {NoStop}%
\bibitem [{\citenamefont {Arrington}\ \emph {et~al.}(2012)\citenamefont
  {Arrington}, \citenamefont {Rubin},\ and\ \citenamefont
  {Melnitchouk}}]{Arrington:2011qt}%
  \BibitemOpen
  \bibfield  {author} {\bibinfo {author} {\bibfnamefont {J.}~\bibnamefont
  {Arrington}}, \bibinfo {author} {\bibfnamefont {J.~G.}\ \bibnamefont
  {Rubin}}, \ and\ \bibinfo {author} {\bibfnamefont {W.}~\bibnamefont
  {Melnitchouk}},\ }\bibfield  {title} {\enquote {\bibinfo {title} {{How Well
  Do We Know The Neutron Structure Function?}}}\ }\href {\doibase
  10.1103/PhysRevLett.108.252001} {\bibfield  {journal} {\bibinfo  {journal}
  {Phys. Rev. Lett.}\ }\textbf {\bibinfo {volume} {108}},\ \bibinfo {pages}
  {252001} (\bibinfo {year} {2012})},\ \Eprint {http://arxiv.org/abs/1110.3362}
  {arXiv:1110.3362 [hep-ph]} \BibitemShut {NoStop}%
\bibitem [{\citenamefont {Airapetian}\ \emph {et~al.}(2011)\citenamefont
  {Airapetian}, \citenamefont {Akopov},\ and\ \citenamefont
  {Akopov}}]{GD11P:2011}%
  \BibitemOpen
  \bibfield  {author} {\bibinfo {author} {\bibfnamefont {A.}~\bibnamefont
  {Airapetian}}, \bibinfo {author} {\bibfnamefont {N.}~\bibnamefont {Akopov}},
  \ and\ \bibinfo {author} {\bibfnamefont {Z.~et~al.}\ \bibnamefont {Akopov}},\
  }\bibfield  {title} {\enquote {\bibinfo {title} {{Inclusive measurements of
  inelastic electron and positron scattering from unpolarized hydrogen and
  deuterium targets}},}\ }\href {\doibase 10.1007/JHEP05(2011)126} {\bibfield
  {journal} {\bibinfo  {journal} {J. High Energ. Phys.}\ }\textbf {\bibinfo
  {volume} {126}} (\bibinfo {year} {2011}),\
  10.1007/JHEP05(2011)126}\BibitemShut {NoStop}%
\bibitem [{\citenamefont {Griffioen}\ \emph {et~al.}(2015)\citenamefont
  {Griffioen} \emph {et~al.}}]{Griffioen:2015hxa}%
  \BibitemOpen
  \bibfield  {author} {\bibinfo {author} {\bibfnamefont {K.~A.}\ \bibnamefont
  {Griffioen}} \emph {et~al.},\ }\bibfield  {title} {\enquote {\bibinfo {title}
  {{Measurement of the EMC Effect in the Deuteron}},}\ }\href {\doibase
  10.1103/PhysRevC.92.015211} {\bibfield  {journal} {\bibinfo  {journal} {Phys.
  Rev.}\ }\textbf {\bibinfo {volume} {C92}},\ \bibinfo {pages} {015211}
  (\bibinfo {year} {2015})},\ \Eprint {http://arxiv.org/abs/1506.00871}
  {arXiv:1506.00871 [hep-ph]} \BibitemShut {NoStop}%
\bibitem [{\citenamefont {Veerasamy}\ and\ \citenamefont
  {Polyzou}(2011)}]{veerasamy11}%
  \BibitemOpen
  \bibfield  {author} {\bibinfo {author} {\bibfnamefont {S.}~\bibnamefont
  {Veerasamy}}\ and\ \bibinfo {author} {\bibfnamefont {W.~N.}\ \bibnamefont
  {Polyzou}},\ }\bibfield  {title} {\enquote {\bibinfo {title} {Momentum-space
  argonne v18 interaction},}\ }\href@noop {} {\bibfield  {journal} {\bibinfo
  {journal} {Phys. Rev. C}\ }\textbf {\bibinfo {volume} {84}},\ \bibinfo
  {pages} {034003} (\bibinfo {year} {2011})}\BibitemShut {NoStop}%
\bibitem [{\citenamefont {Frankfurt}\ and\ \citenamefont
  {Strikman}(1978)}]{FrankfurtStrikman:1978}%
  \BibitemOpen
  \bibfield  {author} {\bibinfo {author} {\bibfnamefont {{L. L.}}\ \bibnamefont
  {Frankfurt}}\ and\ \bibinfo {author} {\bibfnamefont {{M. I.}}\ \bibnamefont
  {Strikman}},\ }\bibfield  {title} {\enquote {\bibinfo {title} {On the problem
  of extracting the neutron structure function from ed scattering},}\ }\href
  {\doibase 10.1016/0370-2693(78)90800-6} {\bibfield  {journal} {\bibinfo
  {journal} {Phys. Lett}\ }\textbf {\bibinfo {volume} {B76}},\ \bibinfo {pages}
  {333--336} (\bibinfo {year} {1978})}\BibitemShut {NoStop}%
\bibitem [{\citenamefont {Wang}\ \emph {et~al.}(2020)\citenamefont {Wang},
  \citenamefont {Thomas},\ and\ \citenamefont {Melnitchouk}}]{Wang:2020uhj}%
  \BibitemOpen
  \bibfield  {author} {\bibinfo {author} {\bibfnamefont {X.G.}\ \bibnamefont
  {Wang}}, \bibinfo {author} {\bibfnamefont {A.W.}\ \bibnamefont {Thomas}}, \
  and\ \bibinfo {author} {\bibfnamefont {W.}~\bibnamefont {Melnitchouk}},\
  }\bibfield  {title} {\enquote {\bibinfo {title} {{Do short-range correlations
  cause the nuclear EMC effect in the deuteron?}}}\ }\href@noop {} {\
  (\bibinfo {year} {2020})},\ \Eprint {http://arxiv.org/abs/2004.03789}
  {arXiv:2004.03789 [hep-ph]} \BibitemShut {NoStop}%
\bibitem [{\citenamefont {Melnitchouk}\ \emph {et~al.}(1997)\citenamefont
  {Melnitchouk}, \citenamefont {Sargsian},\ and\ \citenamefont
  {Strikman}}]{Melnitchouk:1997pemc}%
  \BibitemOpen
  \bibfield  {author} {\bibinfo {author} {\bibfnamefont {W.}~\bibnamefont
  {Melnitchouk}}, \bibinfo {author} {\bibfnamefont {M.}~\bibnamefont
  {Sargsian}}, \ and\ \bibinfo {author} {\bibfnamefont {M.}~\bibnamefont
  {Strikman}},\ }\bibfield  {title} {\enquote {\bibinfo {title} {{Probing the
  origin of the EMC effect via tagged structure functions of the deuteron}},}\
  }\href {\doibase 10.1007/s002180050372} {\bibfield  {journal} {\bibinfo
  {journal} {Z. Phys.}\ }\textbf {\bibinfo {volume} {A359}},\ \bibinfo {pages}
  {99--109} (\bibinfo {year} {1997})}\BibitemShut {NoStop}%
\bibitem [{\citenamefont {Hen}\ \emph {et~al.}(2014)\citenamefont {Hen} \emph
  {et~al.}}]{hen14}%
  \BibitemOpen
  \bibfield  {author} {\bibinfo {author} {\bibfnamefont {O.}~\bibnamefont
  {Hen}} \emph {et~al.},\ }\bibfield  {title} {\enquote {\bibinfo {title}
  {{Momentum sharing in imbalanced Fermi systems}},}\ }\href {\doibase
  10.1126/science.1256785} {\bibfield  {journal} {\bibinfo  {journal}
  {Science}\ }\textbf {\bibinfo {volume} {346}},\ \bibinfo {pages} {614--617}
  (\bibinfo {year} {2014})},\ \Eprint {http://arxiv.org/abs/1412.0138}
  {arXiv:1412.0138 [nucl-ex]} \BibitemShut {NoStop}%
\bibitem [{\citenamefont {Duer}\ \emph {et~al.}(2018)\citenamefont {Duer} \emph
  {et~al.}}]{duer18}%
  \BibitemOpen
  \bibfield  {author} {\bibinfo {author} {\bibfnamefont {M.}~\bibnamefont
  {Duer}} \emph {et~al.} (\bibinfo {collaboration} {CLAS Collaboration}),\
  }\bibfield  {title} {\enquote {\bibinfo {title} {{Probing high-momentum
  protons and neutrons in neutron-rich nuclei}},}\ }\href {\doibase
  10.1038/s41586-018-0400-z} {\bibfield  {journal} {\bibinfo  {journal}
  {Nature}\ }\textbf {\bibinfo {volume} {560}},\ \bibinfo {pages} {617--621}
  (\bibinfo {year} {2018})}\BibitemShut {NoStop}%
\end{thebibliography}%

\end{document}